\documentclass[a4paper,11pt]{article}

\pdfoutput=1
\usepackage{jcappub}
\usepackage{graphicx}
\usepackage{multirow}
\usepackage{hyperref}
\usepackage{amsmath} % Required for \sech
\DeclareMathOperator{\sech}{sech} % Define \sech manually
\usepackage{physics}
\usepackage{float}
\usepackage{amssymb}
\usepackage{graphicx}
\usepackage{bm}
\usepackage{latexsym}
\usepackage{epsfig}
\usepackage{psfrag}
\usepackage{color}
\usepackage[dvipsnames]{xcolor}
\usepackage{subfigure}
\usepackage[utf8]{inputenc}
\usepackage[a4paper, total={7in, 9in}]{geometry}
\DeclareMathOperator{\arcsinh}{arcsinh}
\usepackage[normalem]{ulem}
\makeatletter
\gdef\@fpheader{}
\makeatother

\newcommand{\be}{\begin{equation}}
\newcommand{\ee}{\end{equation}}
\newcommand{\bea}{\begin{eqnarray}}
\newcommand{\eea}{\end{eqnarray}}

\allowdisplaybreaks

\title{Production of Gravitational Waves from Preheating and Tachyonic Instabilities% after Inflation
}
\author[a]{Khursid Alam,}
\author[b]{Koushik Dutta,}
\author[b]{Ahamadullah Khan}

\affiliation[a]{ Department of Physics, Ashoka University, Rajiv Gandhi Education City, Rai, Sonipat, 131029, Haryana, India}
\affiliation[b]{Department of Physical Sciences, Indian Institute of Science 
Education and Research Kolkata, Mohanpur, Nadia 741246, India}
\emailAdd{khursid.alam@ashoka.edu.in}
\emailAdd{ak22rs019@iiserkol.ac.in}
\emailAdd{koushik@iiserkol.ac.in}

\begin{document} 

\begin{abstract}
    {We analyze GW production during preheating for an $\alpha$-attractor potential terminating in the positive-curvature regime, with energy transfer via $\phi\chi^{2}$. Linear Floquet analysis and nonlinear simulations show that $\phi$ fluctuations grow by parametric resonance, while $\chi$ undergoes tachyonic bursts. The GW spectrum features two peaks: a dominant low-frequency peak from the parametric channel and a subdominant high-frequency peak from the tachyonic channel. Redshifted to today, the peak reaches $h^{2}\Omega_{\rm GW}^{(0)} \sim 10^{-11}$ at $f^{(0)}_{p} \sim 10^{7}$ Hz. This multi-peak structure is a characteristic imprint of trilinear preheating in $\alpha$-attractors.}
\end{abstract}
\maketitle

%%%%%%%%%%%%%%%%%%%%%%%%%%%%%%%%%%%%%%%%%%%%%%%%%%%%%%%%%%%%%%%%%%%%%%%%%%%%%%%
\section{Introduction}
\label{Introduction}
The inflationary paradigm~\cite{PhysRevD.23.347, Linde:1981mu} provides a compelling framework for understanding the large--scale homogeneity, isotropy, and flatness of the observed universe, as well as the nearly scale-invariant spectrum of primordial curvature perturbations imprinted in the cosmic microwave background (CMB). A period of accelerated expansion in the very early universe, driven by a scalar field (the inflaton) slowly rolling down a sufficiently flat potential, stretches quantum fluctuations to cosmological scales and seeds the formation of structure. However, inflation must eventually connect to the standard hot Big Bang evolution through a transition phase in which the energy stored in the inflaton condensate is efficiently transferred into a thermal bath of relativistic particles. This process, known as \textit{reheating}~\cite{Kofman:1997, Lozanov:2019, PhysRevD.53.1776, Allahverdi:2010xz}, determines the initial conditions for the radiation-dominated era and can affect inflationary predictions such as the scalar spectral index $n_s$ and the tensor-to-scalar ratio $r$~\cite{Planck:2018jri}.

The dynamics of this transition are often highly non-equilibrium. Instead of proceeding via slow, perturbative decays of the inflaton, the transfer of energy can occur explosively through non-perturbative particle production---a stage known as \textit{preheating}~\cite{Kofman:1997, Lozanov:2019, Kaiser:1997mp, Greene:1997fu, Kaiser:1997hg,Amin:2014eta,Dufaux:2006, Abolhasani:2010}. During preheating, the coherent oscillations of the inflaton about the minimum of its potential lead to the resonant amplification of field fluctuations. Depending on the nature of the inflaton’s couplings, the curvature of the potential and the base-potential of the inflaton field, the fluctuations may grow through either parametric resonance~\cite{Kofman:1997, Lozanov:2019, Kaiser:1997mp, Greene:1997fu, Kaiser:1997hg,Amin:2014eta} or tachyonic instability~\cite{Dufaux:2006, Abolhasani:2010}. The former arises when the time-dependent effective frequency of a field violates the adiabaticity condition which in turn leads to narrow or broad resonance bands; the latter occurs when the effective frequency squared becomes negative for part of each oscillation, leading to exponential growth of the corresponding modes. These nonlinear processes generate large inhomogeneities and mark the first nonlinear stage in the post-inflationary universe.

Among the broad landscape of inflationary models, the $\alpha$--attractor class \cite{Kallosh:2013a,Kallosh:2013b,Linde:2013,Kallosh:2022} offers a particularly robust and well-motivated framework. These models naturally emerge in supergravity and conformal field constructions and predict universal relations among CMB observables that are in excellent agreement with the latest Planck results. The potentials feature an exponentially flat plateau at large field values and a smooth transition to a positively curved minimum where inflation ends. In this regime, the inflaton oscillates coherently about the minimum, providing a natural stage for studying preheating dynamics.

Energy transfer from the inflaton to other degrees of freedom can proceed through several types of couplings.
In particular, trilinear interactions of the form $h\phi\chi^{2}$ induce short, intense bursts of tachyonic amplification in the daughter field $\chi$: as $\phi$ oscillates, the effective frequency-squared of $\chi$ can become negative for part of an oscillation, producing rapid growth of certain $\chi$ modes, known as \textit{tachyonic resonance} \cite{Dufaux:2006, Abolhasani:2010}. Separately, the inflaton field itself can undergo \textit{parametric resonance}~\cite{Kofman:1997, Lozanov:2019, Kaiser:1997mp, Greene:1997fu, Kaiser:1997hg,Amin:2014eta} when its effective frequency becomes time-dependent due to the oscillating inflaton background. Such time dependence can violate the adiabaticity condition for certain $\phi$ modes, leading to exponential amplification of the corresponding fluctuations through broad or narrow resonance. The coexistence and interplay of these two mechanisms—parametric resonance of $\phi$ and tachyonic amplification of $\chi$—can drive efficient energy transfer, strong backreaction, and fragmentation of the inflaton condensate~\cite{Figueroa:2016wxr,Garcia:2023eol,Garcia:2023dyf}.

A remarkable consequence of such nonlinear dynamics can generate a stochastic background of \textit{gravitational waves} (GWs). The inhomogeneous field configurations generated during preheating source tensor perturbations through their anisotropic stresses, producing a stochastic \textit{gravitational waves} background (SGWB) that decouples almost immediately after the end of preheating and free-streams to the present epoch. The spectrum of this SGWB---its shape, peak frequency, and amplitude---encode valuable information about the underlying field dynamics and couplings, offering a unique probe of the post-inflationary universe \cite{Easther:2006gt,GarciaBellido:2007,Figueroa:2017,Cosme:2023}. Depending on the inflationary energy scale, the resulting GW signals can span a wide range of frequencies, often extending into the MHz–GHz range, potentially accessible to future ultra–high–frequency detectors \cite{Aggarwal:2020olq, Aggarwal:2025noe}.

In this work, we investigate the production of gravitational waves during preheating in an $\alpha$--attractor inflationary model where inflation ends in the positive-curvature regime of the potential. We consider a trilinear interaction between the inflaton field $\phi$ and a light scalar daughter field $\chi$, described by the term $h\,\phi\,\chi^2$. Using both linear analysis and nonlinear lattice simulations, we examine how energy is transferred from the inflaton to the daughter field through the combined effects of parametric and tachyonic resonances. In the linear regime, we analyze the characteristic growth rates of the field fluctuations. In the nonlinear regime, we employ the \texttt{CosmoLattice} framework \cite{Figueroa:2021,Figueroa:2023} to solve the full system, including backreaction and rescattering effects.

Our simulations reveal a distinctive two–stage structure in the preheating dynamics. Initially, short bursts of tachyonic amplification in $\chi$ trigger rapid but transient growth of long-wavelength modes. Subsequently, parametric resonance in the inflaton field dominates, leading to a broadband excitation of modes and a turbulent cascade. These amplified inhomogeneities act as strong sources of gravitational radiation. The resulting GW spectrum exhibits a characteristic double–peak structure: a dominant low–frequency peak sourced by the parametric channel and a subdominant high–frequency peak from tachyonic bursts. Redshifted to the present epoch, the spectrum peaks at $f_p \sim 10^{7}\,\mathrm{Hz}$ with an amplitude $h^{2}\Omega_{\mathrm{GW}}^{(0)} \sim 10^{-11}$. This multi–peak feature constitutes a distinctive signature of trilinear preheating in $\alpha$--attractor models.

The remainder of this paper is organized as follows. In Section~\ref{sec:inflation_potential_intial_condition}, we summarize the $\alpha$--attractor potential and derive the initial conditions for preheating. In Section~\ref{sec:potential_during_preheating_phase}, we introduce the trilinear interaction and describe the effective potential structure. Section~\ref{sec:resonances} presents the analysis of both parametric and tachyonic resonances using linear as well as nonlinear lattice simulations. In Section~\ref{sec:GWs_production_during_preheating}, we compute the gravitational wave spectrum generated during preheating and discuss its redshift to the present day. Finally, Section~\ref{sec:summary} summarizes our main results and outlines prospects for future theoretical and observational developments.

%%%%%%%%%%%%%%%%%%%%%%%%%%%%%%%%%%%%%%%%%%%%%%%%%%%%%%%%%%%%%%%%%%%%%%%%%%%%%%%
\section{$\alpha$-attractor inflation and initial conditions for preheating}
\label{sec:inflation_potential_intial_condition}
In this section, we discuss the \(\alpha\)-attractor inflationary potential, which drives inflation. We determine the potential parameters that satisfy CMB constraints, along with the initial field value and velocity of the inflaton field for starting preheating. 

The general form of the \(\alpha\)-attractor potential is given by \cite{Kallosh:2013daa,Kallosh:2013hoa,Kallosh:2013yoa,Kallosh:2022ggf}  
\begin{equation}
    V_{\text{inf}}(\phi) = \frac{\Lambda^4}{p}\,\tanh^p\left(\frac{\phi}{M}\right),
    \label{inflatonpotential}
\end{equation}  
where \(\Lambda\), \(M\), and \(p\) are the potential parameters. The exponent \(p\) is typically an even integer; in our analysis, we set \(p = 4\). One important property of this kind of potential is that the potential is exponentially flat during inflation, but at the end of the inflation, it has the approximate form of $\sim \phi^p$ around its minimum. 

To describe inflationary dynamics, we introduce the first slow-roll parameter, defined as  
\begin{equation}
    \epsilon_{\rm v} = \frac{M_{\rm pl}^2}{2}\left(\frac{V_{\rm inf,\phi}}{V_{\rm inf}}\right)^2,
    \label{slow-roll}
\end{equation}  
where \(V_{\rm inf,\phi}\) denotes the derivative of the potential with respect to the inflaton field and $M_{\rm pl}$ denotes the reduced Planck mass. Inflation proceeds as long as \(\epsilon_{\rm v} < 1\), and it ends when $\epsilon_{\rm v} = 1$. In the regime \(\phi \gg M\), the inflaton field rolls slowly along a flat part of the potential, sustaining inflation. As the field evolves and enters the region \(\phi \ll M\), it begins oscillating around the minimum of the potential, marking the transition to the reheating phase. Solving the condition $\epsilon_{\rm v} = 1$, we express the inflaton field in terms of the potential parameters $M$ and $p$ as
\begin{equation}
    \phi_{*} = \frac{M}{2} \arcsinh\left(\frac{\sqrt{2} p M_{\text{pl}}}{M}\right) \label{inf_value_at_end_of_inflation}.
\end{equation}
This value serves as the initial condition for the inflaton field at the onset of preheating.

\begin{figure}[h!]
    \centering
    \includegraphics[width=0.6\textwidth]{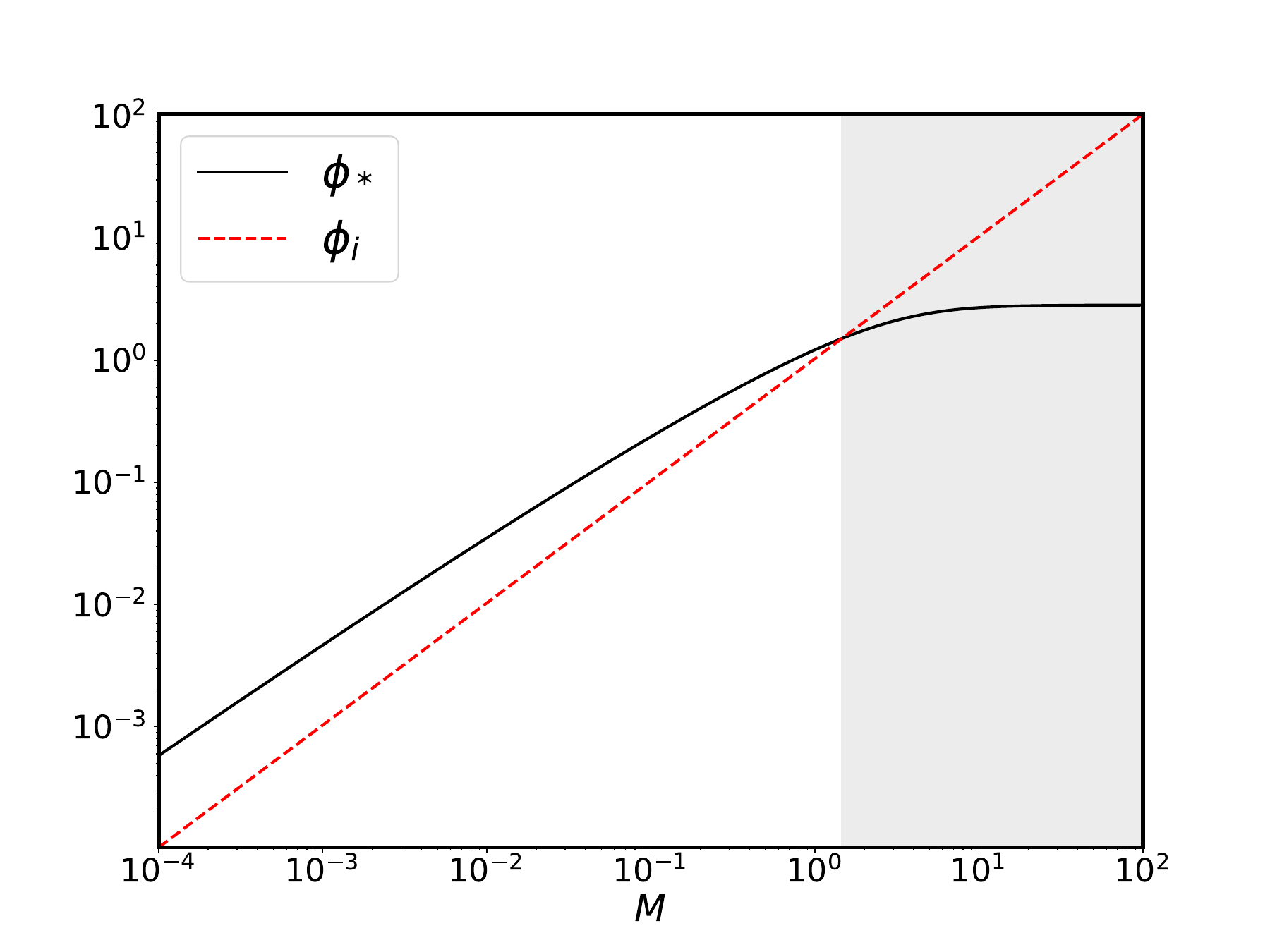}
    \caption{The field value at the inflection point (red dashed line) and the field value at the end of inflation (black solid line) are shown as functions of \( M \) in units of \( M_{\rm pl} = 1 \). The shaded region denotes the domain of positive curvature in the potential.}
    \label{fig:initialfld}
\end{figure}

%%%%%%%%%%%%%%%%%%%%%%%%%%%%%%%%%%%%%%%%%%%%%%%%%%%%%%%%%%%%%%%%%%%%%%%%%%%%%%%
The \(\alpha\)-attractor potential of the inflaton field has an inflection point, which separates the negative and positive curvature regimes of the inflaton potential of Eq.~\eqref{inflatonpotential}. The value of the inflaton field at the inflection point is given by
\begin{equation}
    \phi_{\text{i}} = M \arcsinh\left(\sqrt{\frac{p-1}{2}}\right).
\end{equation}  
For \(\phi \leq \phi_{\text{i}}\), the potential has positive curvature, whereas for \(\phi \geq \phi_{\text{i}}\), the curvature becomes negative. In Fig.~\eqref{fig:initialfld}, the black continuous line represents the end of inflation (that is also the initial inflaton field value for the subsequent preheating phase) for different values of the potential parameter $M$, and the red dashed line indicates the inflaton field value at which the inflection point occurs for various values of the potential parameter $M$. 
We observe that inflation ends at a region of positive curvature in the potential for $M \gtrsim M_{\text{pl}}$, and it is the shaded region in the plot. To avoid tachyonic instability, we choose the initial inflaton field value such that the end of inflation lies within this positively curved region. This ensures that the Bunch-Davies vacuum can safely describe the initial quantum fluctuations. For our analysis, we fix \( M = 10  M_{\rm pl} \).

The homogeneous background dynamics of the inflaton field during inflation are governed by the equation
\begin{equation}
    \ddot{\phi}_{0} + 3H\dot{\phi}_{0} + \frac{\partial V_{\rm inf}}{\partial \phi_{0}}=0, \label{back_ground_dyn_during_inf}  
\end{equation}
where $H = d \ln a / dt$ is the Hubble parameter, and $a$ is the scale factor. The subscript `0' denotes the background inflaton field, which depends only on time, and an overdot represents differentiation with respect to cosmic time. Under the slow-roll approximations ($\ddot{\phi}_{0} \ll 3H\dot{\phi}_{0}$ and $\dot{\phi}^2_{0} \ll V_{\rm inf}$), the above equation in terms of the e-folding number $N$ becomes
\begin{equation}
    \frac{d\phi_{0}}{dN}=-\frac{1}{3H^2}\frac{\partial V_{\rm inf}}{\partial \phi_{0}}~. \label{N_equation}
\end{equation}
The solution of the above equation allows us to 
express the inflaton field at the time when a mode of comoving wavenumber $k$ exits the horizon in terms of the model parameters $p$ and $M$ as
\begin{equation}
    \phi_{k}=\frac{M}{2} \operatorname{arccosh}\left(\cosh{\frac{2 \phi_{*}}{M}} + \frac{4 p M^2_{\rm pl} }{M^2}N_{k}\right), \label{CMB_phi_value}
\end{equation}
where $N_{k}=\ln{a_{\rm *}/a_{k}}$ is the number of e-folds counted from the end of inflation to the horizon exit of the CMB modes.

The scalar power spectrum at horizon exit ($k = aH$)  for a given mode $k$ is given by  
\begin{equation}
    \mathcal{P}_{s}(k)= \frac{1}{12\pi^2}\frac{V_{\rm inf}^3}{M^6_{\rm pl} (V_{\rm inf, \phi})^2} \Bigg|_{k=aH}= \frac{1}{24\pi^2}\frac{2}{p^3} \frac{M^2\Lambda^4}{M^6_{\rm pl}} \frac{\tanh^{p+2}{\left(\frac{\phi_{k}}{M}\right)}}{\sech^4{\left(\frac{\phi_{k}}{M}\right)}}.
\end{equation}
Here $\phi_k$ represents the inflaton field value when the mode $ k $ exits the horizon.
 The Planck 2018 data provides the scalar power spectrum amplitude at the pivot scale $ k_{\rm CMB} = 0.05 $ Mpc$^{-1}$, measured to be $ 2.1\times 10^{-9} $. At the pivot scale, the corresponding inflaton field value, denoted as $ \phi_{\rm CMB} $, is given in Eq.~\eqref{CMB_phi_value}. Substituting $ \phi_k = \phi_{\rm CMB} $ into the expression for $ \mathcal{P}_s(k) $, we can determine the parameter $ \Lambda^4 $ for different values of $ M $. For our choice of parameter $M = 10M_{\rm pl}$, we find $\Lambda^4 = 8\times 10^{-9}$ in the units of  $M_{\rm pl}$.
 
We now compute the velocity of the inflaton field at the end of inflation, which provides the initial condition for our numerical analysis of the preheating phase. Inflation is assumed to end when the slow-roll parameter $\epsilon_{H} \equiv -\dot{H}/H^{2}$ reaches unity. Under this condition, the velocity of inflaton at the end of inflation can be expressed in terms of the potential as  
\begin{equation}
    \dot{\phi}_{*} \simeq -\sqrt{V(\phi_{0})}\,\Big|_{\phi_{0} = \phi_{*}} .
\end{equation}
This relation is then used to set the initial velocity of the inflaton field in our numerical simulations of the preheating dynamics.

%%%%%%%%%%%%%%%%%%%%%%%%%%%%%%%%%%%%%%%%%%%%%%%%%%%%%%%%%%%%%%%%%%%%%%%%%%%%%%%%%%
\section{Model during preheating in the presence of trilinear interaction}
\label{sec:potential_during_preheating_phase}

In this section, we will discuss the potential of a model where $\alpha$-attractor inflation potential is supplemented with trilinear interaction between the inflaton field $\phi$ and a daughter field $\chi$. As discussed above, preheating is the phase that follows the end of inflation, during which the inflaton field oscillates around the minimum of its potential. In the context of $\alpha$-attractor models, this corresponds to the regime $\phi \ll M$, where the slow-roll approximation is no longer valid. Since preheating occurs near the minimum of the potential, and we have chosen $M \gtrsim M_{\rm pl}$, we can approximate the inflaton potential by expanding the Eq.~\eqref{inflatonpotential}. For the case $p = 4$, to leading order, the potential takes the form
\begin{equation}
    V_{\rm inf} \simeq \frac{1}{4} \lambda_{\phi} \phi^4,
\end{equation}
where the effective quartic coupling is given by $\lambda_{\phi} = (\Lambda/M)^4$. We remind ourselves that for our case, we have chosen $M = 10 M_{\rm pl}$, and it subsequently fixed $\Lambda \sim 9.4\times 10^{-3} M_{\rm pl}$, and that leads to $\lambda_{\phi} \sim 8 \times 10^{-13}$. 

To study the transfer of energy from the inflaton to other fields during preheating, we introduce a coupling between the inflaton field $\phi$ and an auxiliary scalar field $\chi$. Specifically, we consider a trilinear interaction of the form $\phi \chi^2$. To make the combined potential bounded from below, we need to add a quartic self-interaction for the $\chi$ field \cite{Dufaux:2006}. Moreover, to ensure that the potential vanishes at its global minimum (barring any contributions for dark energy), we need to add a constant term. In the end, the total potential is given by \cite{Abolhasani:2010}
\begin{equation}
    V(\phi,\chi)= \frac{1}{4}\lambda_{\phi} \phi^4 + \frac{1}{2}h \phi \chi^2 + \frac{1}{4}\lambda_{\chi} \chi^4 + \frac{h^4}{16 \lambda_{\phi}\lambda_{\chi}},
    \label{eq:total_potential}
\end{equation}
where $h$ is the trilinear coupling strength with mass dimension one, and $\lambda_\chi$ denotes the quartic self-coupling of the $\chi$ field.

To perform numerical simulations later, it is convenient to recast the system in terms of dimensionless variables always denoted by $\sim$ . We rescale our relevant quantities as
\begin{equation}  
\omega_* = \sqrt{\lambda_\phi} \phi_*, \quad d \Tilde{\eta} = a^{-1} \, \omega_* dt, \quad d\tilde{x}^{i} = \omega_* dx^{i}, \quad \tilde{f} = \frac{f}{\phi_*}, \quad \mathcal{\tilde{H}} = \frac{H}{a^{-1}\omega_*}, \quad \tilde{V}= \frac{V}{\phi_*^2 \omega_*^2} \label{rescale_equ}.
\end{equation}  
Here, $f$ stands for either $\phi$ or $\chi$, $\tilde{\eta}$ and $\tilde{x}$ are the dimensionless conformal time and spatial coordinate, and $\tilde{\mathcal{H}}$ denotes the dimensionless conformal Hubble parameter.

In terms of these variables, the rescaled dimensionless potential becomes
\begin{equation}
    \tilde{V}(\tilde{\phi}, \tilde{\chi})  
 = \frac{1}{4} \left(\tilde{\phi}^2 - \frac{q_3^2}{2 q_\chi}\right)^2 + \frac{q_\chi}{4}  \left(\tilde{\chi}^2 + \frac{q_3 \tilde{\phi}}{ q_\chi}\right)^2,
    \label{eq:rescaled_potential}
\end{equation} 
where the dimensionless couplings are defined as
\begin{equation}
    q_3 = \frac{h \phi_*}{\omega_*^2}, \quad q_\chi = \frac{\lambda_\chi \phi_*^2}{\omega_*^2}. \label{coupling_const}
\end{equation}
The potential has a global minimum at
\begin{equation}
    \tilde{\phi}_{min} = -\frac{q_3}{\sqrt{2 q_\chi}}, \quad \tilde{\chi}_{min}^2 = - \frac{q_3 \tilde{\phi}_{min}}{ q_\chi}.
    \label{eq:global_minima}
\end{equation}
To ensure that the constant term in the potential does not affect the inflationary phase, we impose the condition \cite{Abolhasani:2010}
\begin{equation} 
    q_\chi \gg \frac{q_3^2}{\tilde{\phi}_{*}^2}.\label{parameters_values}
\end{equation}
This requirement guarantees that inflation proceeds as usual and ensures numerical stability in simulations. For all our chosen values of $q_3$, the constant term in Eq.~\eqref{eq:total_potential} is set such that the above condition is satisfied, which corresponds to a value of approximately $10^{-14}$.

\begin{figure}[H]
    \centering
    \includegraphics[width=4.5in, height = 3.6in]{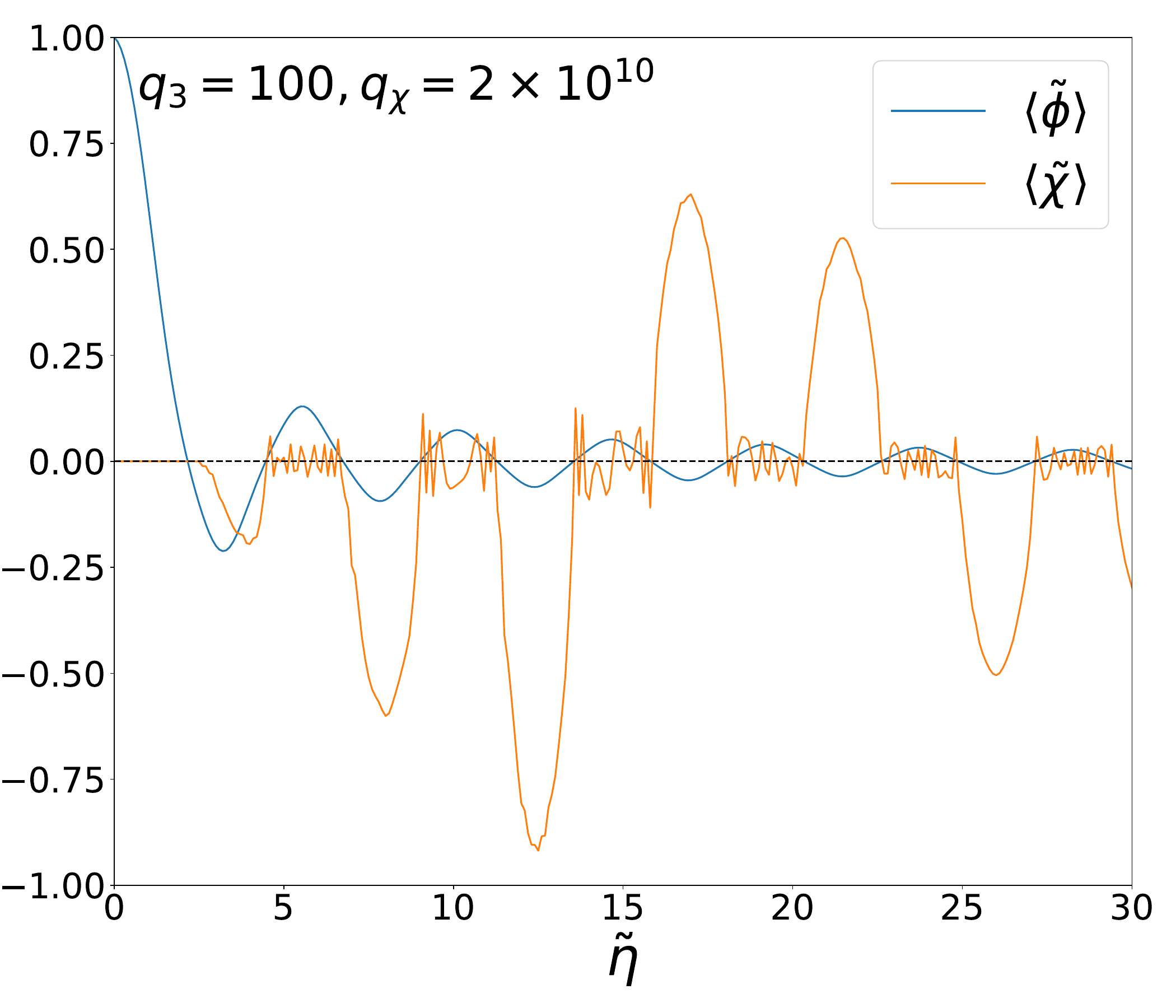}
    \caption{The time evolution of the spatially averaged inflaton field, \( \langle \tilde{\phi} \rangle \), and the daughter field, \( \langle \tilde{\chi} \rangle \), is shown.}
    \label{fig:phi_chi_comp}
\end{figure}

To understand how this potential influences the dynamics, we write the equations of motion of both inflaton field $\tilde \phi$ and the daughter field $\tilde \chi$ as
\begin{align}
    &\Tilde{\phi}'' - \frac{\Tilde{\nabla}^2}{a^2} \Tilde{\phi} + 2 \Tilde{\mathcal{H}} \Tilde{\phi}' + a^2 \Tilde{\phi}^3 + \frac{q_3}{2} a^2 \Tilde{\chi}^2 = 0, \label{rescalephi_eq}\\
    &\Tilde{\chi}'' - \frac{\Tilde{\nabla}^2}{a^2} \Tilde{\chi} + 2 \Tilde{\mathcal{H}} \Tilde{\chi}' + a^2 q_3 \Tilde{\phi} \Tilde{\chi} + a^2 q_\chi \Tilde{\chi}^3 = 0, \label{rescalechi_eq}
\end{align}
where primes denote derivatives with respect to $\tilde \eta$. We use \texttt{CosmoLattice} code~\cite{Figueroa:2021,Figueroa:2023} to solve the above equations numerically. Although the above equations are written for space-time dependent fields, in this section we confine ourselves to the time-dependent parts of both $\tilde \phi$ and $\tilde \chi$. This can be achieved by taking the spatial average of these fields. Time evolution of the inflaton and the daughter field is shown in Fig.~\ref{fig:phi_chi_comp}.

Analyzing the structure of the potential given in Eq.~\eqref{eq:rescaled_potential}, we find that the inflaton field $\tilde{\phi}$ has a single global minimum located at $\tilde{\phi}_{\min}$. After inflation ends, the inflaton oscillates around the above mentioned minimum during the preheating phase. In contrast, the dynamics of the auxiliary field $\tilde{\chi}$ are more complex. The potential along the $\tilde{\chi}$ direction admits two extrema, which occur at
\begin{equation}
    \tilde{\chi} = 0, \quad \text{and} \quad \tilde{\chi}^2 = -\frac{q_3 \tilde{\phi}}{q_\chi}.
\end{equation}

At the end of inflation the $\tilde \chi$ sits at  $\tilde \chi = 0$ whereas $\tilde \phi > 0$. As the $\tilde \phi$ moves towards its global minimum, and eventually becomes negative the minimum at $\tilde \chi =0$ becomes a maximum for the $\tilde \chi$ direction, and the new minimum is shifted to $\tilde{\chi} = \pm \sqrt{\frac{q_3 |\tilde{\phi}|}{q_\chi}}$. In this stage, the $\tilde \chi$ field moves in a negatively curved potential. Subsequently, when $\tilde \phi$ becomes positive again, $\tilde \chi$ starts oscillating around its then minimum $\tilde \chi =0$. Note that movement of the $\tilde \chi$ field towards its doubly degenerate minima depends on the field velocity at the local maxima. This is depicted by the orange color in Fig.~\ref{fig:phi_chi_comp}. We will see in the next section that during the movement of the $\tilde \chi$ field along the negatively curved potential, the effective mass squared of the fluctuations for the field becomes negative for a short period of time, triggering a tachyonic instability. %This leads to a rapid amplification of $\tilde{\chi}$ modes, a process known as tachyonic resonance.

\section{Analysis of parametric and tachyonic resonances}
\label{sec:resonances}
In this section, we discuss the behavior of the fluctuations of both the $\tilde{\phi}$ and $\tilde{\chi}$ fields. To begin with, we first study the linearized versions of Eqs.~\eqref{rescalephi_eq} and~\eqref{rescalechi_eq}, which in momentum space take the form
\begin{align}
    &\delta\Tilde{\phi}_{\tilde{k}}'' + 2\Tilde{\mathcal{H}}\, \delta\Tilde{\phi}_{\tilde{k}}' + \left[\tilde{k}^2 + 3 a^2 \Tilde{\phi}_0^2\right] \delta\Tilde{\phi}_{\tilde{k}} = 0, \label{phi_k_eq}\\
    &\Tilde{\chi}_{\tilde{k}}'' + 2\Tilde{\mathcal{H}}\, \Tilde{\chi}_{\tilde{k}}' + \left[\tilde{k}^2 + a^2 q_3 \Tilde{\phi}_0 + 3a^2 q_4 \langle\Tilde{\chi}^2\rangle \right] \Tilde{\chi}_{\tilde{k}} = 0. \label{chi_k_eq}
\end{align}
Here, \( \tilde{k} = k/\omega_{*} \) is the dimensionless comoving wavenumber, and $\tilde{\phi}_0 = \langle \tilde{\phi} \rangle$ denotes the background inflaton field. In writing Eq.~\eqref{chi_k_eq}, we have used the Hartree approximation $\tilde{\chi}^3  \rightarrow 3\tilde{\chi} \langle \tilde{\chi}^2 \rangle$ to linearize it. Moreover, in $\Tilde{\mathcal{H}}$, we have neglected the contributions from field fluctuations. 

The effective frequencies for the inflaton and daughter field modes are then given by
\begin{align}
    &\tilde{\omega}^2_{\tilde{k}, \tilde{\phi}} = \tilde{k}^2 + 3 a^2 \Tilde{\phi}_0^2, \label{freq_phi}\\
    &\tilde{\omega}^2_{\tilde{k}, \tilde{\chi}} = \tilde{k}^2 + a^2 q_3 \Tilde{\phi}_0 + 3 a^2 q_4 \langle \Tilde{\chi}^2\rangle \equiv \tilde{k}^2 + \tilde{m}^2_{\tilde{\chi}, \mathrm{eff}}, \label{freq_chi}
\end{align}
where \( \tilde{m}^2_{\tilde{\chi}, \mathrm{eff}} \) denotes the effective mass squared of the daughter field, sourced by both the background inflaton field and the Hartree correction arising from the self-interaction of the $\tilde{\chi}$ field.

\begin{figure}[h!]
\centering
\includegraphics[width=3.4in, height = 2.6 in]{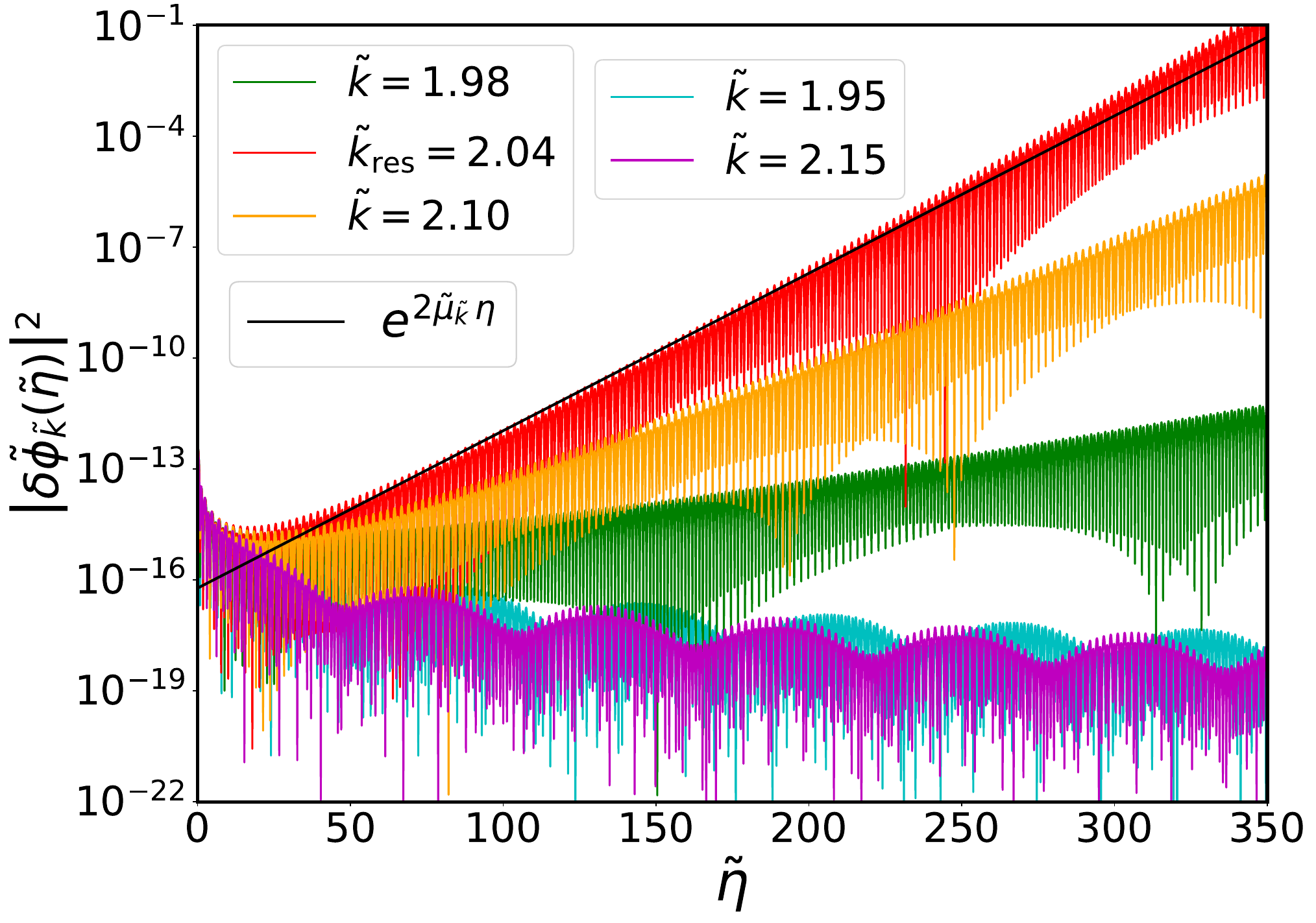}
\hspace{0.2cm}
\includegraphics[width=3.4in, height = 2.6 in]{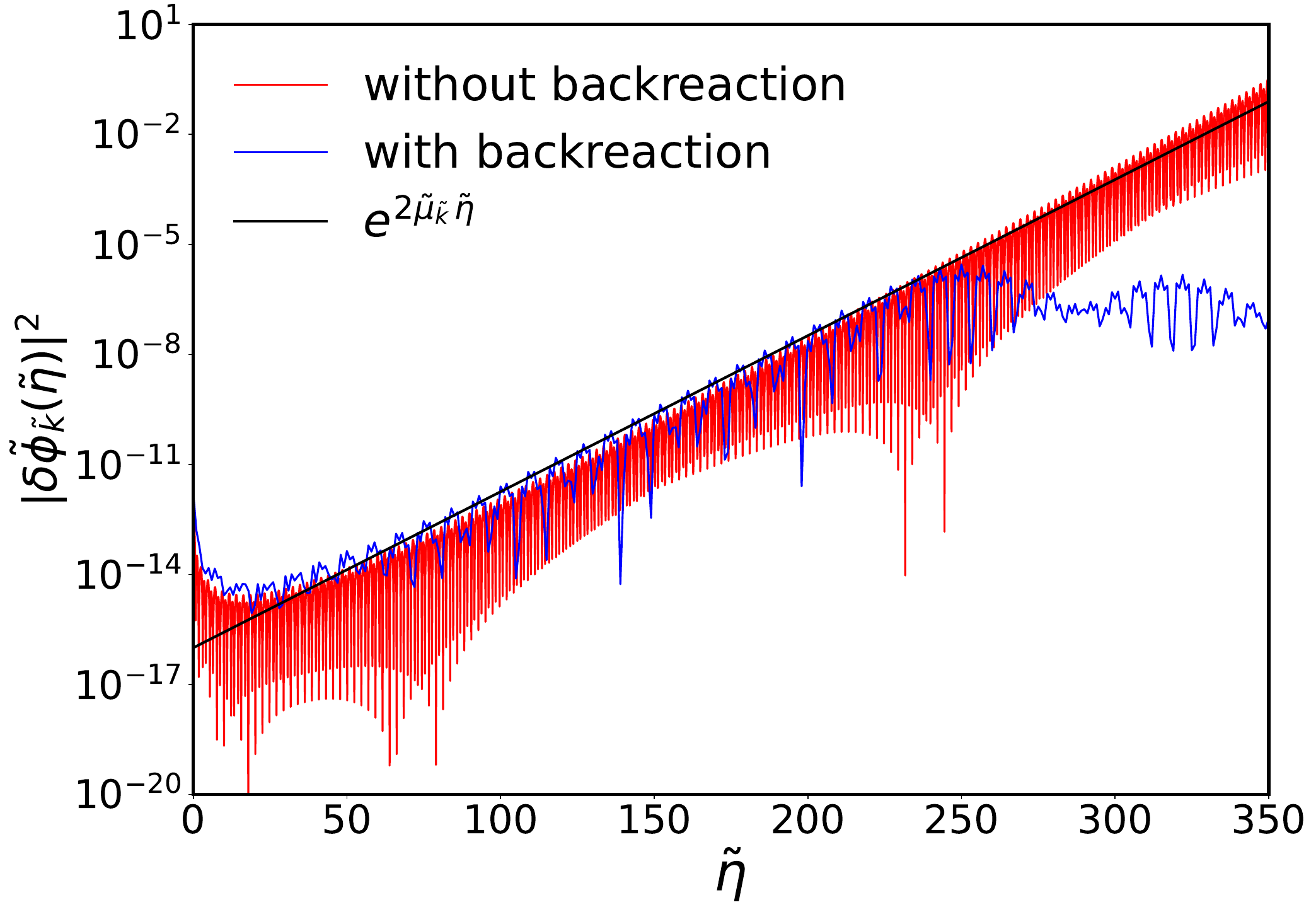}
\vskip -0.1in
\caption{In the left panel, the time evolution of the inflaton field fluctuation, \( \delta \tilde{\phi}_{\tilde{k}} \), for different modes \( \tilde{k} \) is depicted by solving Eq.~\eqref{phi_k_eq}. In the right panel, the evolution of \( \delta \tilde{\phi}_{\tilde{k}} \) for the resonant mode \( \tilde{k}_{\rm res} = 2.04 \) is displayed: (i) obtained by solving Eq.~\eqref{phi_k_eq} (red line) and (ii) computed using \texttt{CosmoLattice} (blue line).
}
\label{fig:preheating_amp}
\end{figure}

First, we examine the evolution of the inflaton field fluctuations, $\delta\tilde{\phi}_{\tilde{k}}$, for various comoving modes $\tilde{k}$, as shown in Fig.~\ref{fig:preheating_amp}. To do this, we solve the mode equation for $\delta\tilde{\phi}_{\tilde{k}}$, given in Eq.~\eqref{phi_k_eq}. The middle term in this equation represents Hubble damping, which acts to dilute the energy stored in $\delta\tilde{\phi}_{\tilde{k}}$ through redshift. The main dynamics, however, are governed by the effective frequency $\tilde{\omega}_{\tilde{k}, \tilde{\phi}}$, which appears in the final term within the square brackets of Eq.~\eqref{phi_k_eq} and is defined in Eq.~\eqref{freq_phi}. This effective frequency is always positive but time dependent. As a result, it can satisfy the \textit{non-adiabaticity condition}, $\tilde{\omega}'_{\tilde{k},\tilde{\phi}}/\tilde{\omega}^2_{\tilde{k},\tilde{\phi}} \gtrsim 1$, for a certain range of $\tilde{k}$ modes. This condition is typically met when the background field $\tilde{\phi}_0$ passes through the minimum of its potential, where the effective mass of the inflaton fluctuations becomes small and the modes can be excited more easily. When this happens, the fluctuations grow exponentially, and a substantial amount of energy is transferred from the coherent oscillations of the background inflaton field into these fluctuation modes. This process is known as parametric resonance and has been widely studied in the literature~\cite{Kofman:1997, Lozanov:2019, Kaiser:1997mp, Greene:1997fu, Kaiser:1997hg, Amin:2014eta}.

During parametric resonance, the fluctuations grow approximately as $e^{\tilde{\mu}_{\tilde{k}}\tilde{\eta}}$, where $\tilde{\mu}_{\tilde{k}}$ is the \textit{Floquet exponent}. A positive value of $\tilde{\mu}_{\tilde{k}}$ signifies an exponential amplification of the corresponding mode. As shown in the left panel of Fig.~\ref{fig:preheating_amp}, only a narrow band of modes, $1.98 \lesssim \tilde{k} \lesssim 2.10$, undergoes significant amplification. Among these, one particular mode, highlighted in red, experiences the strongest growth. We refer to this mode as the resonant mode, denoted by $\tilde{k}_{\rm res}$. In our setup, the resonant mode occurs at $\tilde{k}_{\rm res} \approx 2.04$, where the growth of $\delta\tilde{\phi}_{\tilde{k}}$ is well described by a Floquet exponent $\tilde{\mu}_{\tilde{k}} \approx 0.049$. The corresponding solution, proportional to $e^{\tilde{\mu}_{\tilde{k}}\tilde{\eta}}$, is shown by the black curve in the left panel of Fig.~\ref{fig:preheating_amp}.

To assess the validity of the linear analysis, we compare its results with the fully nonlinear numerical solutions obtained using lattice simulations, as shown in the right panel of Fig.~\ref{fig:preheating_amp}. In this figure, the solid red line represents the inflaton fluctuations obtained by solving Eq.~\eqref{phi_k_eq}, while the solid blue line shows the corresponding results from the lattice simulation for the resonant mode, $\tilde{k}_{\rm res}$. In the linear analysis, the fluctuations grow exponentially forever because of the neglect of both backreaction and nonlinear effects. In contrast, the lattice simulation includes these effects, leading to a slowdown and eventual saturation of the growth once the fluctuations become sufficiently large. At early times, when the fluctuation amplitudes are still small, both approaches agree closely, showing that the linear approximation works well in this regime. However, as the system evolves and nonlinearities become important, the linear analysis fails, and only the lattice simulation can capture the correct dynamics. For this reason, in the rest of our analysis, we rely on \texttt{CosmoLattice} to solve the system self-consistently, fully accounting for backreaction and nonlinear terms.

We now study the time evolution of the fluctuations of the daughter field for a given comoving mode $\tilde{k}$. For this purpose, similar to $\delta\tilde{\phi}_{\tilde{k}}$, we consider the linearized equation of motion for the Fourier modes of the daughter field, given by Eq.~\eqref{chi_k_eq}. This equation governs the evolution of each mode under a time-dependent effective mass, and consequently, a time-dependent frequency, as given in Eq.~\eqref{freq_chi}. Typically, when the inflaton field takes sufficiently negative values, the effective mass squared becomes negative, i.e., $\tilde{m}^2_{\tilde{\chi},\mathrm{eff}} < 0$. From Eq.~\eqref{freq_chi}, the effective frequency squared for a given mode $\tilde{k}$, which satisfies $\tilde{k} < |\tilde{m}_{\tilde{\chi},\mathrm{eff}}|$ for $\tilde{m}^2_{\tilde{\chi},\mathrm{eff}} < 0$, becomes negative. Hence, the daughter field fluctuations corresponding to these modes can grow exponentially, a process known as tachyonic resonance~\cite{Dufaux:2006, Abolhasani:2010}. This mechanism plays the main role in transferring energy from the inflaton to the daughter field during the early stages of preheating.

\begin{figure}[h]
\centering
\includegraphics[width=3.2in, height = 2.4 in]{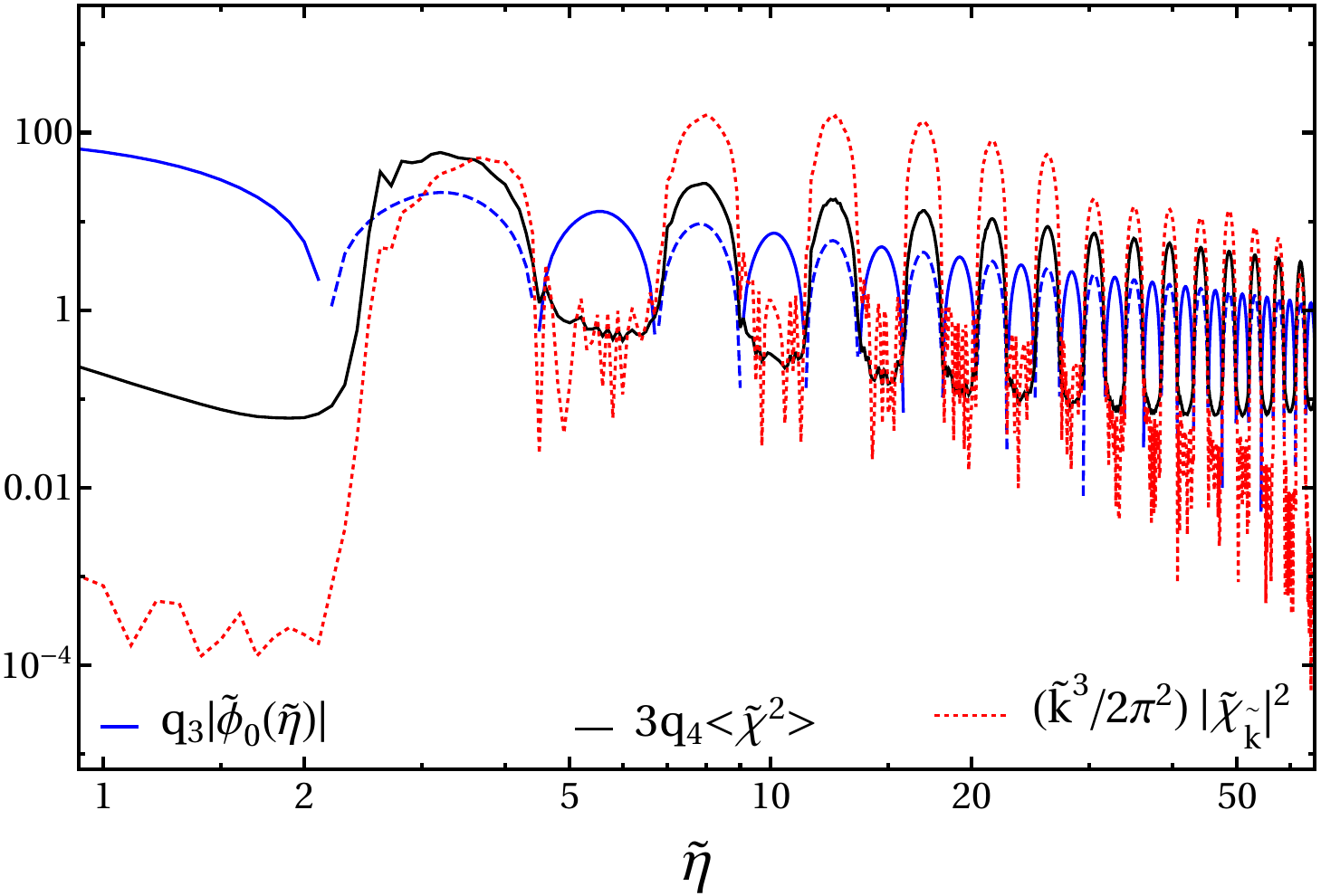}
\hspace{0.2cm}
\includegraphics[width=3.2in, height = 2.4
in]{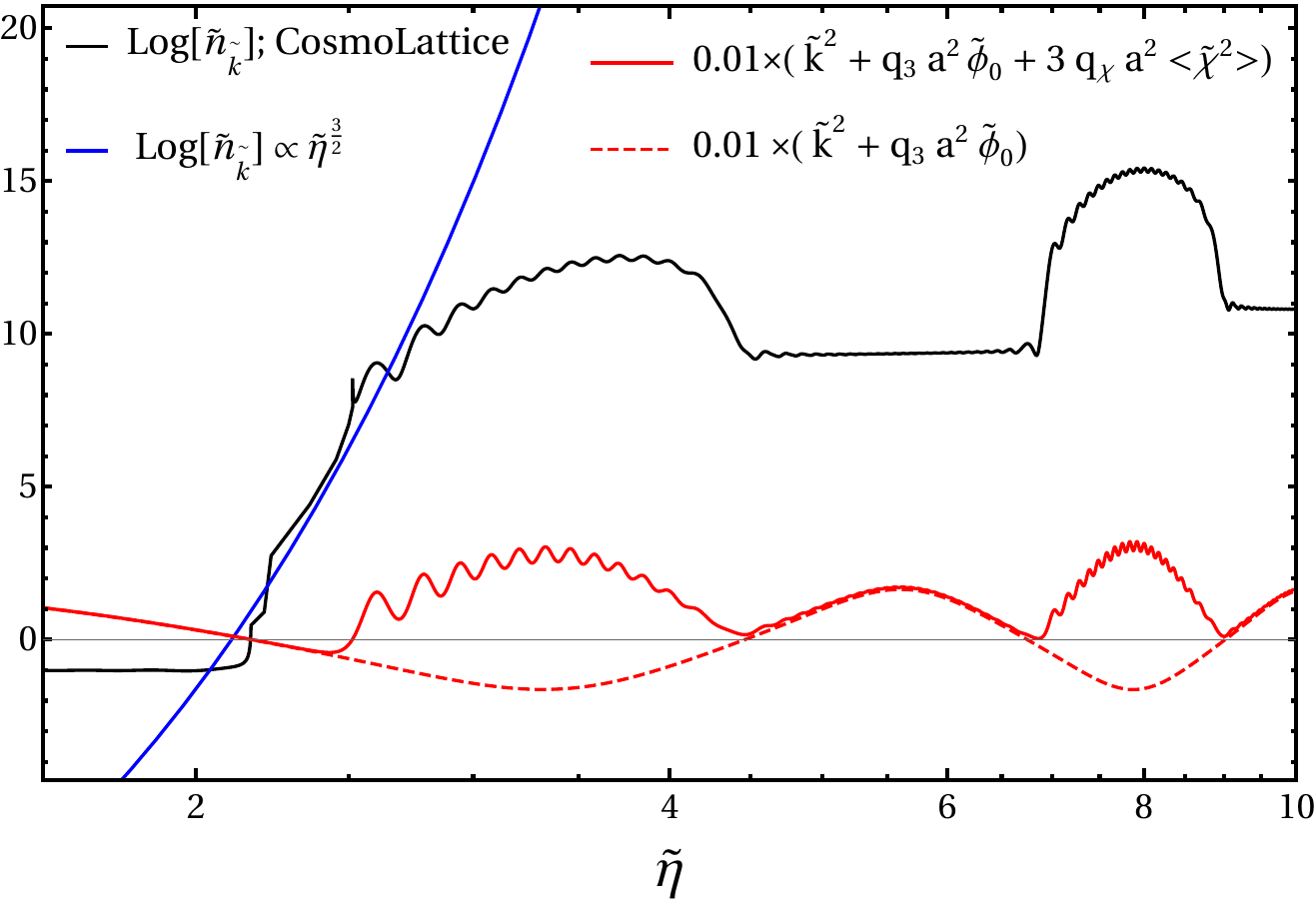}
\vskip -0.1in
\caption{In the left panel, the time evolution of \( q_3 \tilde{\phi}_0 \) (blue) and \( 3 q_4 \langle \tilde{\chi}^2 \rangle \) (black) is depicted. The blue dashed segments indicate intervals during which the inflaton field \( \tilde{\phi}_0 \) takes negative values. The growth of the fluctuation for a given mode \( \tilde{k} \) is illustrated by the red dashed line. In the right panel, the blue and black curves represent the occupation number of the daughter field obtained analytically, \( n_{\tilde{k}, \tilde{\chi}} \propto e^{\tilde{\eta}^{3/2}} \)~\cite{Abolhasani:2010}, and from lattice simulations, respectively. The red solid and red dashed curves denote the effective frequency \( \tilde{\omega}_{\tilde{k}, \tilde{\chi}}^2 \) with and without backreaction effects, respectively. For both panels, the parameters \( \tilde{k} = 1.2 \), \( q_3 = 100 \), and \( q_{\chi} = 2 \times 10^{10} \) are used.
}

\label{fig:cond_for_tachyonic_growth}
\end{figure}

The left panel of Fig.~\ref{fig:cond_for_tachyonic_growth} illustrates the onset of tachyonic instability during the early stages of preheating. The blue curve shows the evolution of $q_3 \tilde{\phi}_{0}(\tilde{\eta})$, where solid segments correspond to positive values of the inflaton field, and dashed segments to negative values. The solid black curve represents the evolution of $3q_4\langle \tilde{\chi}^2 \rangle$, which captures the variance of the daughter field, while the red dashed curve displays the power spectrum of a specific comoving mode $\tilde{k}$ of the daughter field. The power spectrum for a given mode $\tilde{k}$ rises sharply over a very short interval, around $\tilde{\eta} \approx 3$, in the regime where $\tilde{\phi}_0(\tilde{\eta}) < 0$, and the contribution from the daughter field is negligible. Together, these satisfy the condition $\tilde{k} < |\tilde{m}_{\tilde{\chi},\mathrm{eff}}|$, or equivalently, $\tilde{\omega}_{\tilde{k}}^2 < 0$. As mentioned earlier, this growth is due to tachyonic resonance induced by the oscillating background inflaton field. Once the term $3q_4\langle \tilde{\chi}^2 \rangle$ dominates over the term $q_3 \tilde{\phi}_{0}(\tilde{\eta})$, the effective frequency squared is no longer negative. This marks the onset of the backreaction time and halts the further growth of the daughter field fluctuations. The corresponding maximum value of $\tilde{\chi}$ is 
\begin{equation}
    \langle \tilde{\chi}^2 \rangle_{\mathrm{max}} \approx \frac{q_3 \tilde{\phi}_0}{3q_4}.
\end{equation}

In the right panel of Fig.~\ref{fig:cond_for_tachyonic_growth}, we plot the effective frequency, $\tilde{\omega}_{\tilde{k}, \tilde{\chi}}^2$, for a representative mode $\tilde{k}$ (shown in red). The red solid curve corresponds to the effective frequency including the backreaction effect, while the red dashed curve represents the case without it. The backreaction can be switched off by dropping the $3q_4\langle \tilde{\chi}^2 \rangle$ term in Eq.~\eqref{chi_k_eq}. In that case, the effective mass squared, $\tilde{m}_{\tilde{\chi},\mathrm{eff}}^2 = a^2 q_3 \tilde{\phi}_0$, exhibits both negative and positive cycles. Consequently, modes satisfying $\tilde{k} < |\tilde{m}_{\tilde{\chi},\mathrm{eff}}|$ correspond to $\tilde{\omega}_{\tilde{k}, \tilde{\chi}}^2 < 0$ whenever the inflaton field $\tilde{\phi}_0$ is in its negative phase (see the red dashed curve in Fig.~\ref{fig:cond_for_tachyonic_growth}). Hence, these modes undergo amplification during such intervals. In contrast, when the backreaction is included, $\tilde{\omega}_{\tilde{k}, \tilde{\chi}}^2 < 0$ holds only for a short duration before turning positive, as shown by the red solid curve in Fig.~\ref{fig:cond_for_tachyonic_growth}.

The growth of the $\tilde{\chi}$ fluctuations can be quantified through the occupation number, $n_{\tilde{k}, \tilde{\chi}}$, defined as  
\begin{equation}
    n_{\tilde{k}, \tilde{\chi}} = \frac{\tilde{\omega}_{\tilde{k}, \tilde{\chi}}}{2} \left( \frac{|\dot{\tilde{\chi}}_{\tilde{k}}|^2}{\tilde{\omega}_{\tilde{k}, \tilde{\chi}}^2} + |\tilde{\chi}_{\tilde{k}}|^2 \right) - \frac{1}{2}.
    \label{occ_numb}
\end{equation}
When the backreaction is neglected, the occupation number grows indefinitely, approximately following the analytical relation $n_{\tilde{k}, \tilde{\chi}} \propto e^{\tilde{\eta}^{3/2}}$, as discussed in Ref.~\cite{Abolhasani:2010}. The right panel of Fig.~\ref{fig:cond_for_tachyonic_growth} shows the numerically computed occupation number, including backreaction (black curve), along with the analytical fit $e^{\tilde{\eta}^{3/2}}$ (blue curve). In our case, the occupation number increases rapidly around $\tilde{\eta} \approx 3$ and then saturates due to the onset of backreaction. During the growth phase, we observe excellent agreement with the analytical behavior $e^{\tilde{\eta}^{3/2}}$, confirming the accuracy of our numerical results. For detailed analytical treatments of tachyonic resonance, see Refs.~\cite{Dufaux:2006} (for the $m^2\phi^2$ model) and \cite{Abolhasani:2010} (for the same model as ours, but without backreaction).

So far, we have seen that fluctuations grow with time for a given mode $\tilde{k}$ depending on certain conditions. To examine how these fluctuations vary with $\tilde{k}$, we calculate the power spectrum for both fields. The ensemble average of the product of the Fourier modes of the fields leads to the power spectrum, $\tilde{\Delta}_{\tilde{f}}(\tilde{k})$, defined as  
\begin{equation}
\tilde{\Delta}_{\tilde{f}}(\tilde{k}) \equiv \frac{\tilde{k}^3}{2\pi^2} \tilde{\mathcal{P}}_{\tilde{f}}(\tilde{k}), \hspace{0.7cm} 
\langle \delta\tilde{f}_{\tilde{k}} \, \delta\tilde{f}_{\tilde{k}'} \rangle \equiv (2\pi)^3 \tilde{\mathcal{P}}_{\tilde{f}}(\tilde{k})\, \delta(\tilde{k} - \tilde{k}').
\label{power_spectrum}
\end{equation}  
The fluctuations of the inflaton field, $\delta\tilde{\phi}_{\tilde{k}}$, grow due to parametric resonance, whereas those of the daughter field, $\tilde{\chi}_{\tilde{k}}$, grow due to tachyonic resonance. This behavior is reflected in the power spectrum, as supported by its definition.

\begin{figure}[h]
\centering
\includegraphics[width=3.4in, height = 2.8 in]{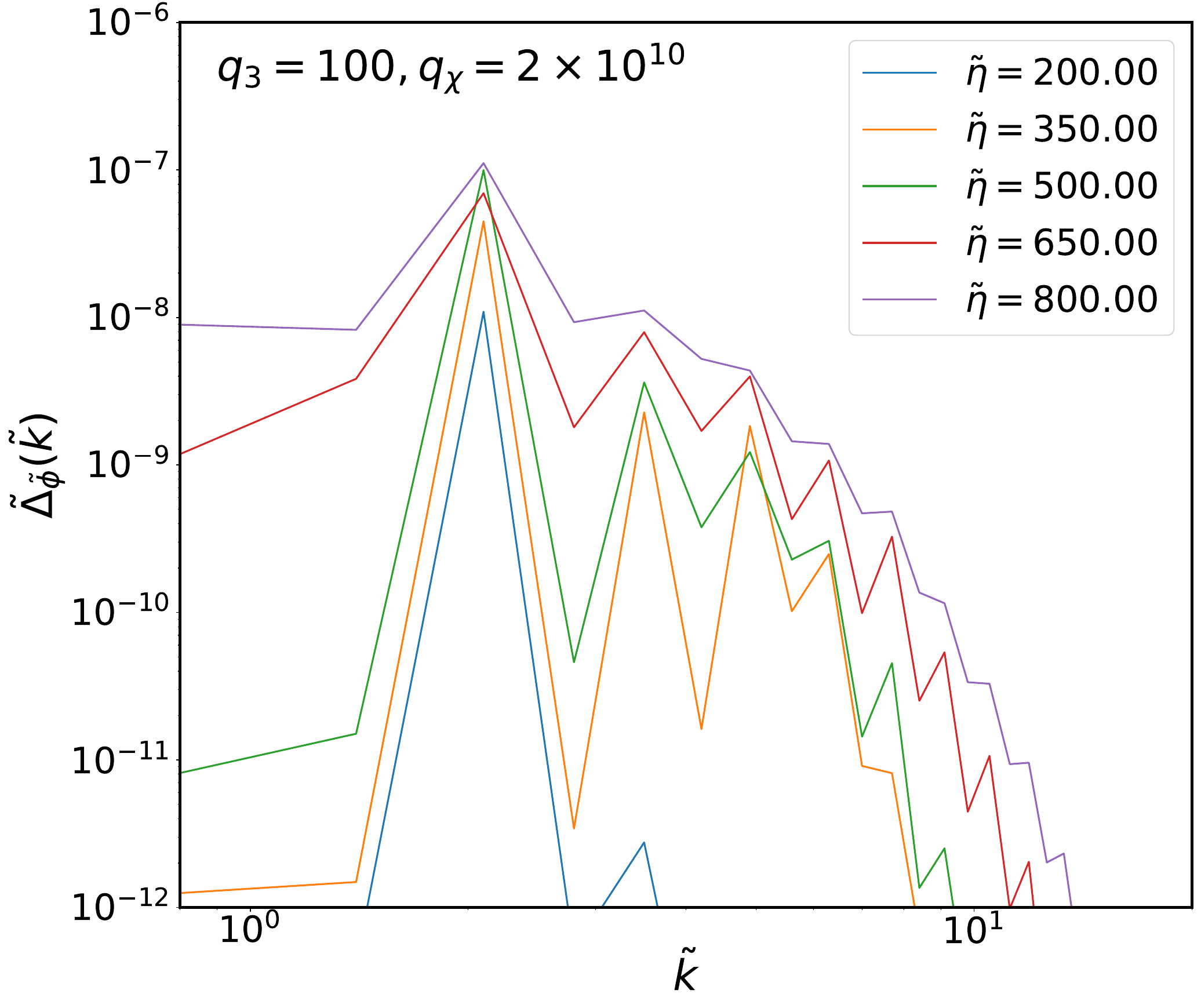}
\hspace{0.2cm}
\includegraphics[width=3.4in, height = 2.8 in]{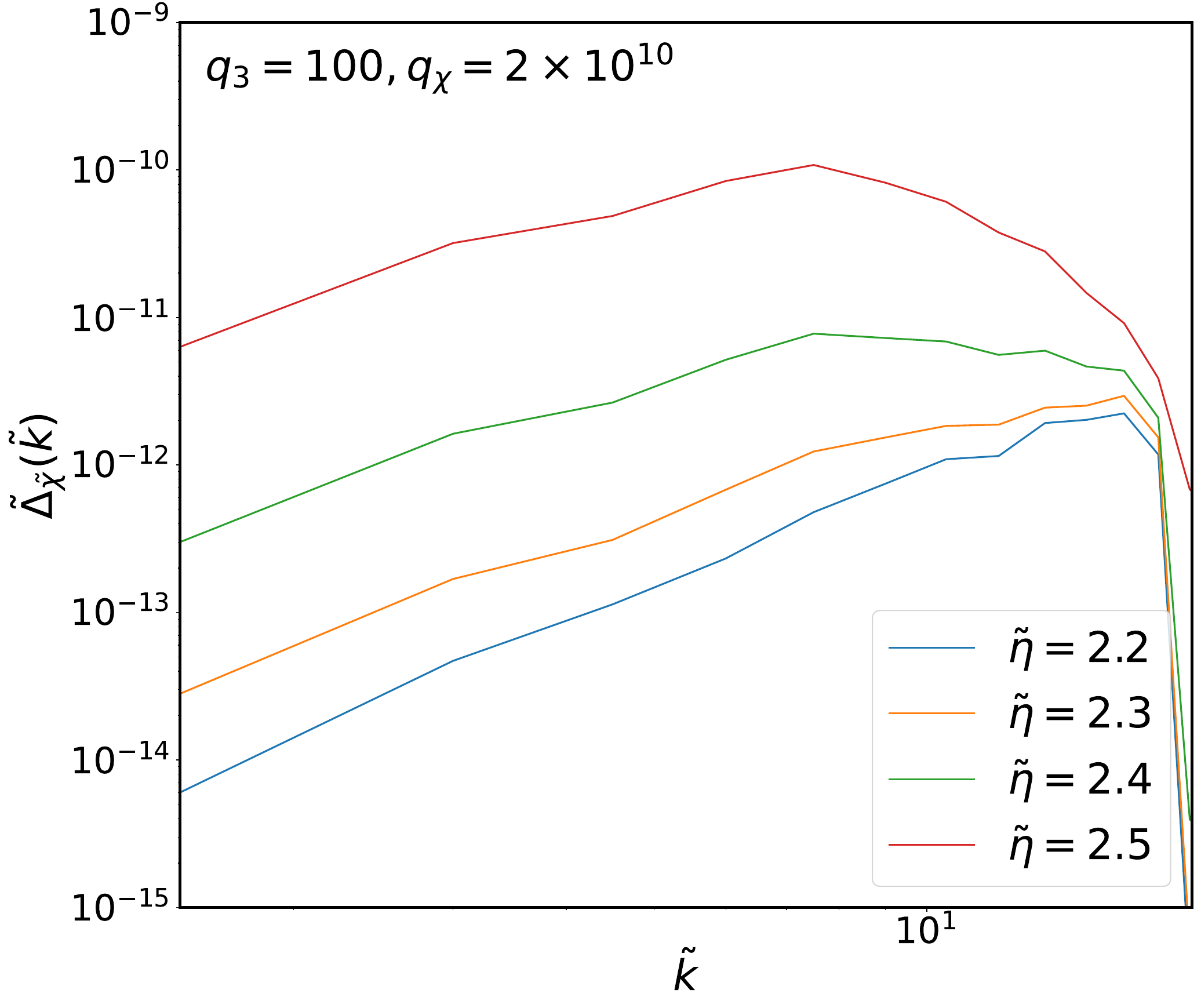}
\vskip -0.1in
\caption{In the left panel, the power spectrum of the inflaton field \( \tilde{\phi} \) is displayed. In the right panel, the power spectrum of the daughter field \( \tilde{\chi} \) is shown.
}
\label{fig:inflaton_daughter_fld_power_sp}
\end{figure}

In the left panel of Fig.~\ref{fig:inflaton_daughter_fld_power_sp}, we plot the power spectrum for the fluctuations of the inflaton field at different times. From the figure, it is clear that as time progresses, $\tilde{\Delta}_{\tilde{\phi}}(\tilde{k})$ increases within a certain range of $\tilde{k}$ around the resonant mode, $\tilde{k}_{\rm res}$, and reaches its maximum at the resonant mode. 
On the other hand, in the right panel of Fig.~\ref{fig:inflaton_daughter_fld_power_sp}, we plot the power spectrum for the daughter field fluctuations at different times. It is also evident from the figure that the fluctuations grow up to certain values of $\tilde{k}$. This behavior entirely depends on $\tilde{m}_{\tilde{\chi},\mathrm{eff}}^2$, defined in Eq.~\eqref{freq_chi}. The maximum value of the mode is then given by $\tilde{k}_{\rm max} = |\tilde{m}_{\tilde{\chi},\mathrm{eff}}^{\rm max}|$; beyond this value, there will be no further growth, as the effective frequency squared, $\tilde{\omega}_{\tilde{k}, \tilde{\chi}}^2$, remains positive thereafter.

\begin{figure}[h]
\centering
\includegraphics[width=3.4in, height = 2.8in]{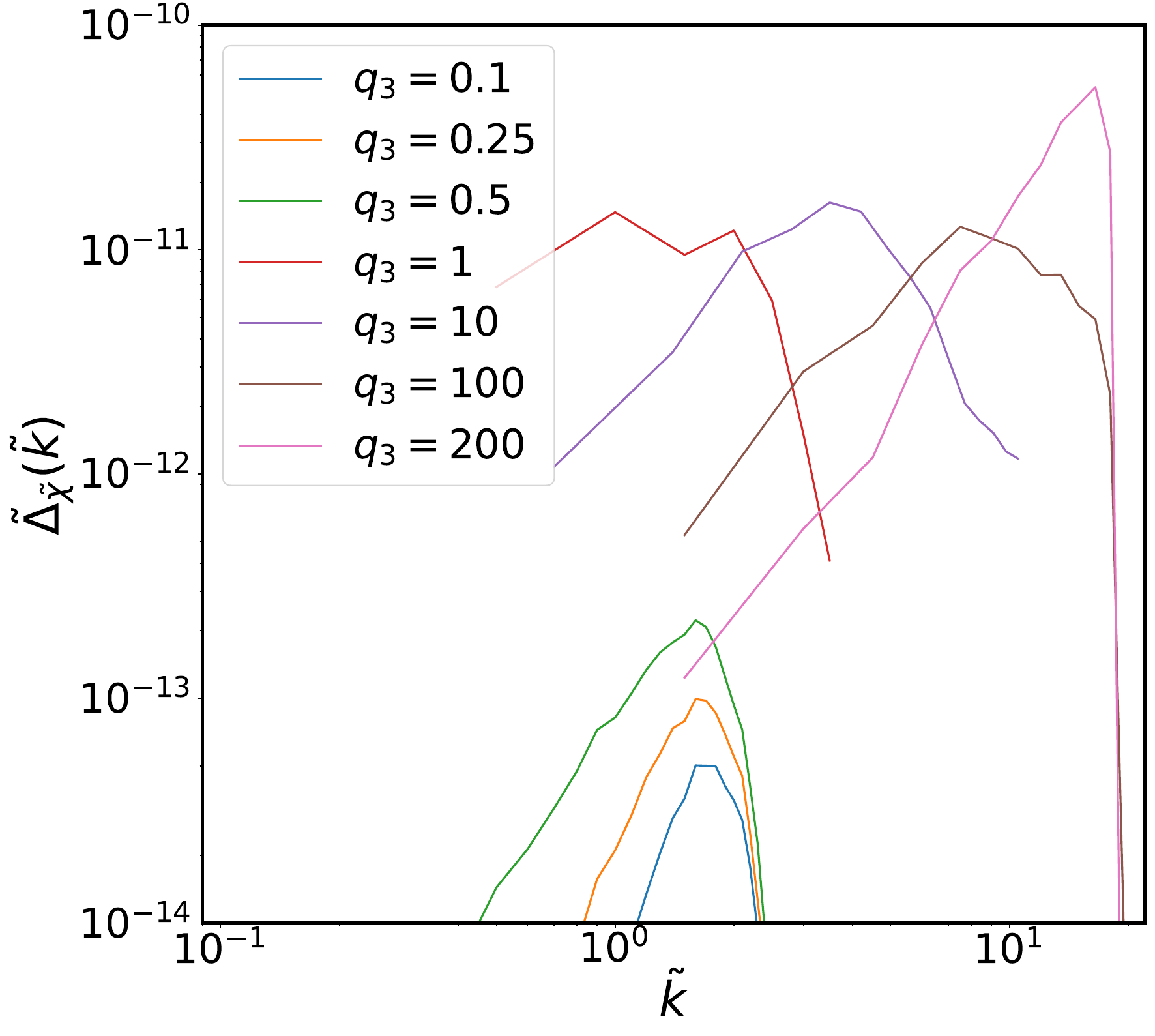}
\hspace{0.05cm}
\includegraphics[width=3.4in, height = 2.8in]{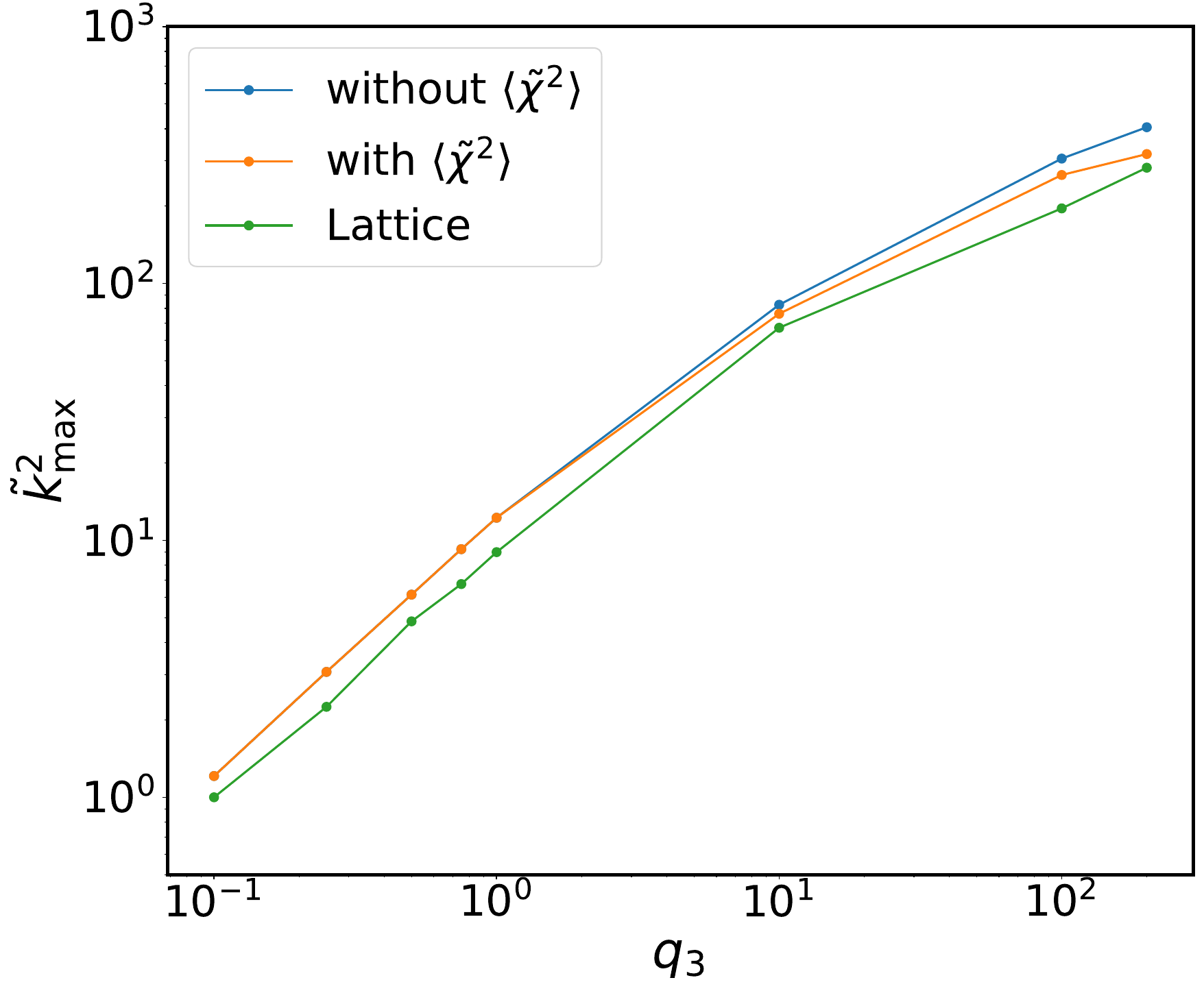}
\vskip -0.1in
\caption{In the left panel, the power spectrum is shown as a function of \( \tilde{k} \) for different values of \( q_3 \), evaluated at the time when \( \tilde{m}_{\tilde{\chi},\mathrm{eff}}^2 \) attains its maximum negative value. In the right panel, the squared value of the mode corresponding to the spectral peak is plotted as a function of \( q_3 \). The blue curve represents \( \tilde{k}_{\mathrm{max}}^2 \) obtained without the \( \langle \tilde{\chi}^2 \rangle \) term in Eq.~\eqref{freq_chi}, while the orange curve includes it. The green curve corresponds to the result extracted from the lattice simulation shown in the left panel.
}
\label{fig:pwsp}
\end{figure}

The production of the daughter field, $\tilde{\chi}$, solely depends on the resonance parameter, $q_3$. This dependence translates to the power spectrum as well as to the modes associated with it. To see this dependence, in the left panel of Fig.~\ref{fig:pwsp}, we show the power spectrum of the daughter field for different values of \( q_3 \). The spectra are evaluated at the time when \( \tilde{m}_{\tilde{\chi},\mathrm{eff}}^2 \) reaches its maximum negative value. We choose this time because the tachyonic instability is strongest then, and modes with momenta smaller than or close to the most unstable mode are efficiently amplified. Therefore, the peak of each curve corresponds to the maximum value of the unstable mode, $\tilde{k}_{\mathrm{max}}$. From the figure, we see that the peaks of the power spectrum increase and move toward higher \( \tilde{k} \)-values as \( q_3 \) increases. This occurs because a larger \( q_3 \) strengthens the effective coupling between the inflaton and the daughter field, allowing energy to be transferred more efficiently from the inflaton condensate. As a result, higher-momentum modes are excited, producing the observed shift in the spectral peak.

The above dependence between \( \tilde{k}_{\max} \) and \( q_3 \) can also be derived with two other approaches, as shown in the right panel of Fig.~\ref{fig:pwsp}. In the first approach, we compute \( \tilde{k}_{\rm max} \) using Eq.~\eqref{freq_chi}, neglecting the contribution of the \( 3q_4\langle\tilde{\chi}^2\rangle \) term in \( \tilde m_{\tilde\chi,\rm eff}^2 \), as shown by the blue curve. In the second approach, we include the contribution of \( 3q_4\langle\tilde{\chi}^2\rangle \), obtained from lattice simulations, in the \( \tilde m_{\tilde\chi,\rm eff}^2 \) term, represented by the orange curve. The two curves agree closely for small \( q_3 \), where the neglected term \( 3q_4\langle\tilde{\chi}^2\rangle \) remains subdominant. Finally, in the third approach, we determine \( \tilde{k}_{\rm max} \) directly from the power spectra shown in the left panel of Fig.~\ref{fig:pwsp}, represented by the green curve. The close agreement among all three methods highlights the robustness of our analysis, with small deviations at larger \( q_3 \) reflecting the increasing importance of the higher-order contribution \( 3q_4\langle\tilde{\chi}^2\rangle \) as mode production becomes more efficient.

\begin{figure}[H]
    \centering
    \includegraphics[width=3.8in, height = 2.6in]{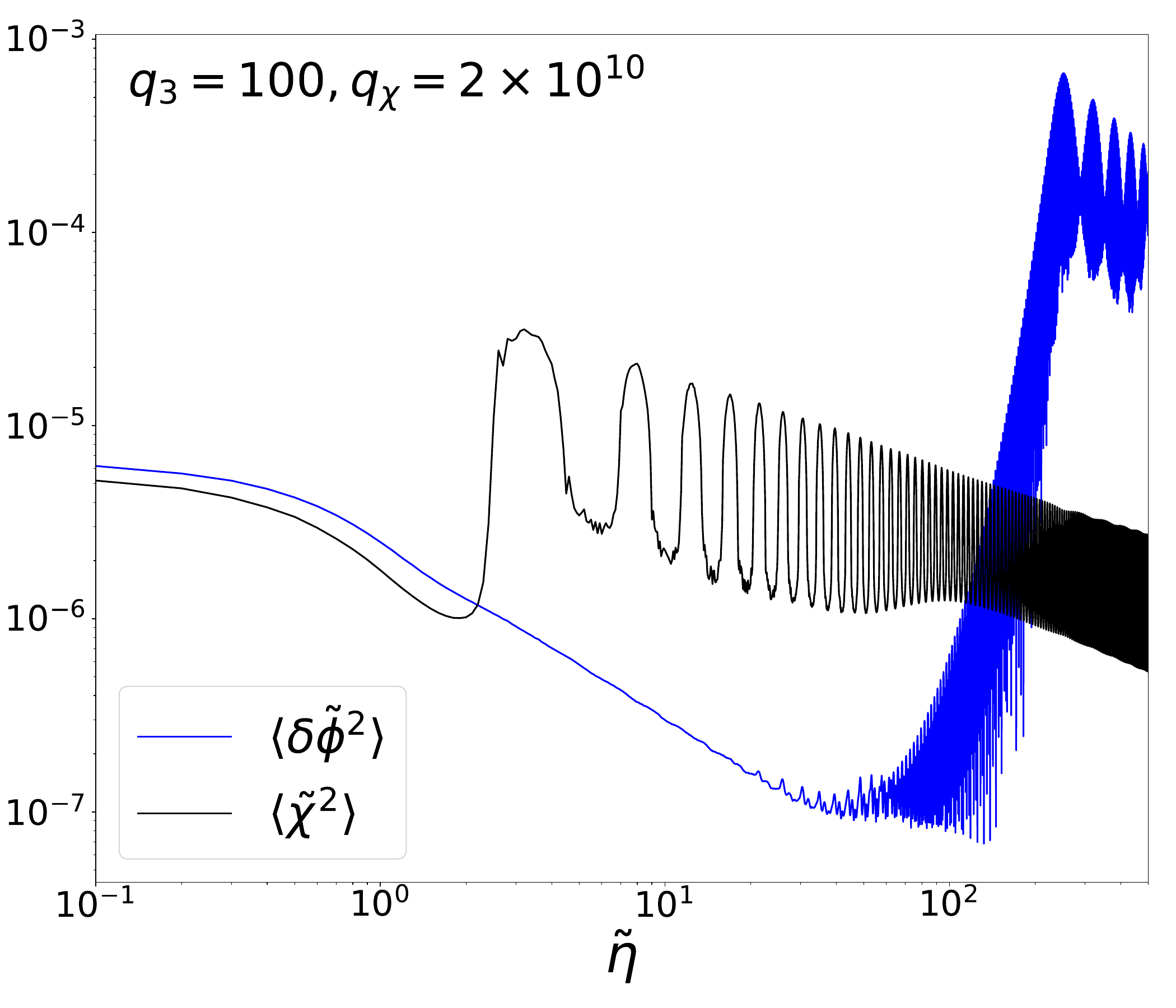}
    \caption{The time evolution of the variance of the inflaton field \( \tilde{\phi} \) and the daughter field \( \tilde{\chi} \) is shown.}
    \label{fig:variance_plot}
\end{figure} 
To conclude this section, we estimate the total growth of fluctuations in both fields by considering all relevant modes. For this purpose, we compute their variances. The general expression for the variance of a field fluctuation in momentum space is given by
\begin{equation}
    \langle \delta \tilde{f}^2 \rangle = \int \frac{d\tilde{k}}{\tilde{k}} \frac{\tilde{k}^3}{2\pi^2} |\delta \tilde{f}_{\tilde{k}}|^2.
\end{equation}
Since we have already obtained the time evolution of the mode functions, $\delta \tilde{\phi}_{\tilde{k}}$ and $\tilde{\chi}_{\tilde{k}}$, the corresponding variances can be evaluated by integrating their spectra over comoving momentum $\tilde{k}$. In practice, these quantities are computed directly as part of the lattice-simulation output.

Fig.~\ref{fig:variance_plot} shows the time evolution of the variances for both fields. The solid blue line corresponds to the inflaton variance, while the solid black line represents that of the daughter field. As evident from the figure, the inflaton variance remains significantly larger than that of the daughter field throughout the evolution. This difference arises because the inflaton fluctuations are enhanced by broad parametric resonance, whereas the daughter-field fluctuations grow mainly through tachyonic instability, which operates only for short intervals. We thus conclude that, in our model, parametric resonance is the dominant channel for transferring energy into field fluctuations compared to the tachyonic resonance.

\section{Production of gravitational waves during preheating}
\label{sec:GWs_production_during_preheating}
In this section, we study the generation of gravitational waves (GWs) during preheating for our model. The production of gravitational waves during preheating has been widely discussed in \cite{Easther:2006gt, Garcia-Bellido:2007nns, Dufaux:2007pt, Dufaux:2008dn, Dufaux:2010cf, Bethke:2013aba, Bethke:2013vca, Figueroa:2017, Adshead:2018doq, Adshead:2019lbr, Cosme:2023}. During this stage, for our case, the fluctuations in both the inflaton and the daughter field grow rapidly: the inflaton undergoes parametric resonance, while the daughter field experiences tachyonic instability. These fluctuations generate large inhomogeneities in the energy-momentum tensor, which subsequently perturb the homogeneous FLRW background metric. 

The transverse and traceless (TT) part of the inhomogeneous energy–momentum tensor generated during preheating acts as a source for gravitational waves. The time evolution of the corresponding metric perturbations, $h_{ij}(\bm{x}, t)$, in the presence of such a source is governed by  
\begin{equation}
    \ddot{h}_{ij}(\bm{x}, t) + 3H\dot{h}_{ij}(\bm{x}, t) - \frac{\nabla^2 h_{ij}(\bm{x}, t)}{a^2} = 16\pi G\,\Pi^{\rm TT}_{ij} (\bm{x},t), \label{GWs_evoluation_equation}
\end{equation}
where $\Pi^{\rm TT}_{ij}(\bm{x}, t)$ denotes the TT projection of the anisotropic stress tensor. This tensor encodes the effect of field inhomogeneities on the GW background and is determined by the spatial gradients of the scalar fields. For scalar fields, the TT part of the anisotropic stress tensor takes the following form
\begin{equation}
    \Pi^{\rm TT}_{ij} = \left\{ \partial_{i}\phi\,\partial_{j}\phi + \partial_{i}\chi\,\partial_{j}\chi \right\}^{\rm TT}.\label{anistropic_stress_tensor}
\end{equation}
These gradient terms act as the main source for GW production during the nonlinear stage of preheating, when both inflaton and daughter field fluctuations become highly amplified and generate significant anisotropic stresses.

To perform numerical simulations, it is convenient to recast the system in terms of dimensionless variables. We rescale the metric perturbation \( h_{ij}(x, t) \) and the energy density \( \rho \) as  
\begin{equation}  
    \tilde{h}_{ij}= h_{ij} \left(\frac{M_{\rm pl}}{\phi_{*}}\right)^2, 
    \qquad 
    \tilde{\rho}= \frac{\rho}{\omega_{*}^2 \phi_{*}^2}.
    \label{gws_rescale_equ}
\end{equation}  
The equation of motion for the Fourier mode \( \tilde{h}_{ij}(\tilde{k}, \tilde{\eta}) \) of the metric perturbation is then given by  
\begin{equation}
    \tilde{h}''_{ij}(\tilde{k}, \tilde{\eta}) 
    + 2 \tilde{\mathcal{H}}\,\tilde{h}'_{ij}(\tilde{k}, \tilde{\eta}) 
    + \tilde{k}^2 \tilde{h}_{ij}(\tilde{k}, \tilde{\eta})
    = 2 \tilde{\Pi}^{TT}_{ij}(\tilde{k}, \tilde{\eta}),
\end{equation}
where \( \tilde{\Pi}^{TT}_{ij}(\tilde{k}, \tilde{\eta}) \) is the Fourier transform of Eq.~\eqref{anistropic_stress_tensor}, rewritten in terms of the dimensionless variables.

The energy density of gravitational waves can be expressed as  
\begin{equation}
    \tilde{\rho}_{\rm GW}(\tilde{\eta}) 
    = \frac{\phi_*^2}{4 a^2 M_{\rm pl}^2} 
      \sum_{i,j} \big\langle \tilde{h}'_{ij}{}^2 \big\rangle_{\tilde{V}}
    = \frac{\tilde{k}_{\rm IR}^3 \phi_*^2}{32 \pi^3 a^2 M_{\rm pl}^2}
      \int \frac{d^3\tilde{k}}{(2\pi)^3} 
      \tilde{h}'_{ij}(\tilde{k}, \tilde{\eta}) \,
      \tilde{h}'^{*}_{ij}(\tilde{k}, \tilde{\eta}),
\end{equation}

where \( \tilde{k}_{\rm IR} \) denotes the minimum infrared (IR) cutoff used in the CosmoLattice simulations. In our analysis, we set \( \tilde{k}_{\rm IR} \) in the range \( [0.7,\,0.8] \).

The abundance of the gravitational waves can be calculated using
\begin{equation}
    h^2 \Omega_{\rm GW}(\tilde k, \tilde \eta)= \frac{h^2}{\tilde \rho_{c}}\frac{d\tilde \rho_{\rm GW}(\tilde k, \tilde \eta)}{d\ln(\tilde k)}= \frac{h^2}{\tilde{\rho}_c} \frac{\tilde{k}^3 \, \tilde{k}^3_{IR}}{4(2\pi)^6 a^2}\left(\frac{\phi_*}{M_{\rm pl}}\right)^2 \int d\Omega_{\tilde k}\tilde h'_{ij}(\tilde k, \tilde \eta)\,\tilde h'^{*}_{ij}(\tilde k, \tilde \eta) ,
\end{equation}
where $\tilde\rho_{c}=\rho_{c}/\omega^2_{*}\phi^2_{*}=3M^2_{\rm pl}\tilde{H}^2/\phi^2_{*}$ represents the critical energy density.

Immediately after the end of preheating, GWs decouple and travel freely throughout space. To observe their signal today, we simply need to redshift the frequency and amplitude appropriately, accounting for the expansion of the universe. The observed frequency today for a comoving mode $k$ produced at the scale factor $a$ is given by 
\begin{equation}
    f_0 = \frac{a}{a_0} \frac{k}{2\pi}.
    \label{freq_today}
\end{equation}
The ratio of the scale factor at the time of production to the scale factor today is given by
\begin{equation}
    \frac{a}{a_0} =\left( \frac{a}{a_{\rm f}} \right)^{\frac{1- 3w}{4}} \left( \frac{a_{\rm f}}{a_{\rm RD}} \right)^{\frac{1- 3\bar{w}}{4}} \left(\frac{g_{s, \rm RD}}{g_{s,0}}\right)^{-\frac{1}{3}} \left(\frac{g_{\rm RD}}{g_0}\right)^{\frac{1}{4}}  \left(\frac{\rho_{0}}{\rho}\right)^{\frac{1}{4}}.
    \label{scale_factor_ratio}
\end{equation} In deriving the above equation, we have used that during the radiation-dominated phase, the energy density ($\rho$) scales with temperature ($T$) as $\rho \propto g_t\, T^4$, and the scale factor and temperature are related by $a\, T \propto g_{s,t}^{-1/3}$, where $g_t$ and $g_{s,t}$ are the effective relativistic and entropic degrees of freedom at time $t$, respectively. We have assumed that the average equation of state parameters from the initial moment $(a_{\rm i})$ to the end of preheating $(a_{\rm f})$, and from the end of preheating to the onset of radiation domination $(a_{\rm RD})$, are $w$ and $\bar{w}$, respectively. We have also assumed that from the onset of radiation domination until today, the universe is dominated by radiation. Finally, the relations between the energy density and the scale factor, given by $\rho \propto a^{-3(1 + w)}$ and $\rho \propto a^{-3(1 + \bar{w})}$ in the two regimes, are also used. Using Eq.~\eqref{scale_factor_ratio}, we can write Eq.~\eqref{freq_today} as

\begin{equation}
          f_0 \approx \left(\frac{\tilde k}{a \,\tilde \rho ^{1/4}}\right)\times 9.86\times 10^6\,   \rm  Hz.
         \label{freq_today_final}  
\end{equation} 
%To get the above equation, we assumed that $w=\bar{\omega}=1/3$.

Another quantity that characterizes GWs is their energy density parameter, $\Omega_{\rm GW}$. At time $t_{\rm f}$, the production of GWs ceases, and the amplitude reaches its maximum value, denoted by $\Omega_{\rm GW}^{(\rm f)}$. The amplitude of the GWs observed today is given by \begin{equation}\begin{aligned}
        h^2\Omega_{\rm GW}^{(0)} & = h^2 \Omega_{\rm RD}^{(0)} \left(\frac{a_{\rm f}}{a_{\rm RD}}\right)^{1- 3\bar{w}} \left(\frac{g_{s, \rm 0}}{g_{s,\rm RD}}\right)^{\frac{4}{3}} \left(\frac{g_{\rm RD}}{g_0}\right) \Omega_{\rm GW}^{(\rm f)}\\
        & \approx \mathcal{O}(10^{-6})\times  4\times \Omega_{\rm GW}^{(\rm f)}.
    \label{amp_today}
\end{aligned}
\end{equation} 
In both Eq.~\eqref{freq_today_final} and Eq.~\eqref{amp_today}, we have used the fact that, since our base potential is $V(\phi) \propto \phi^4$, it corresponds to $w = 1/3$. After preheating ends, the energy density is dominated by the light daughter field $\tilde{\chi}$, which also leads to $\bar{w} = 1/3$. During the radiation-dominated (RD) era, we can reasonably approximate $g_{s,t} \approx g_t$. The ratio we use is $g_0/g_{\rm RD} \approx 100$, which gives $\left( g_0/g_{\rm RD} \right)^{1/12} \sim \mathcal{O}(1)$ and $\left( g_0/g_{\rm RD} \right)^{1/3} \sim \mathcal{O}(0.1)$. Finally, we used the energy density of relativistic species today, $\rho_0 \approx 5.39 \times 10^{-15}~\mathrm{eV}^4$, and $h^2\Omega_{\rm RD}^{(0)} \approx 4.16\times 10^{-5}$.

\begin{figure}[h]
\centering
\includegraphics[width=3.4in, height = 2.8 in]{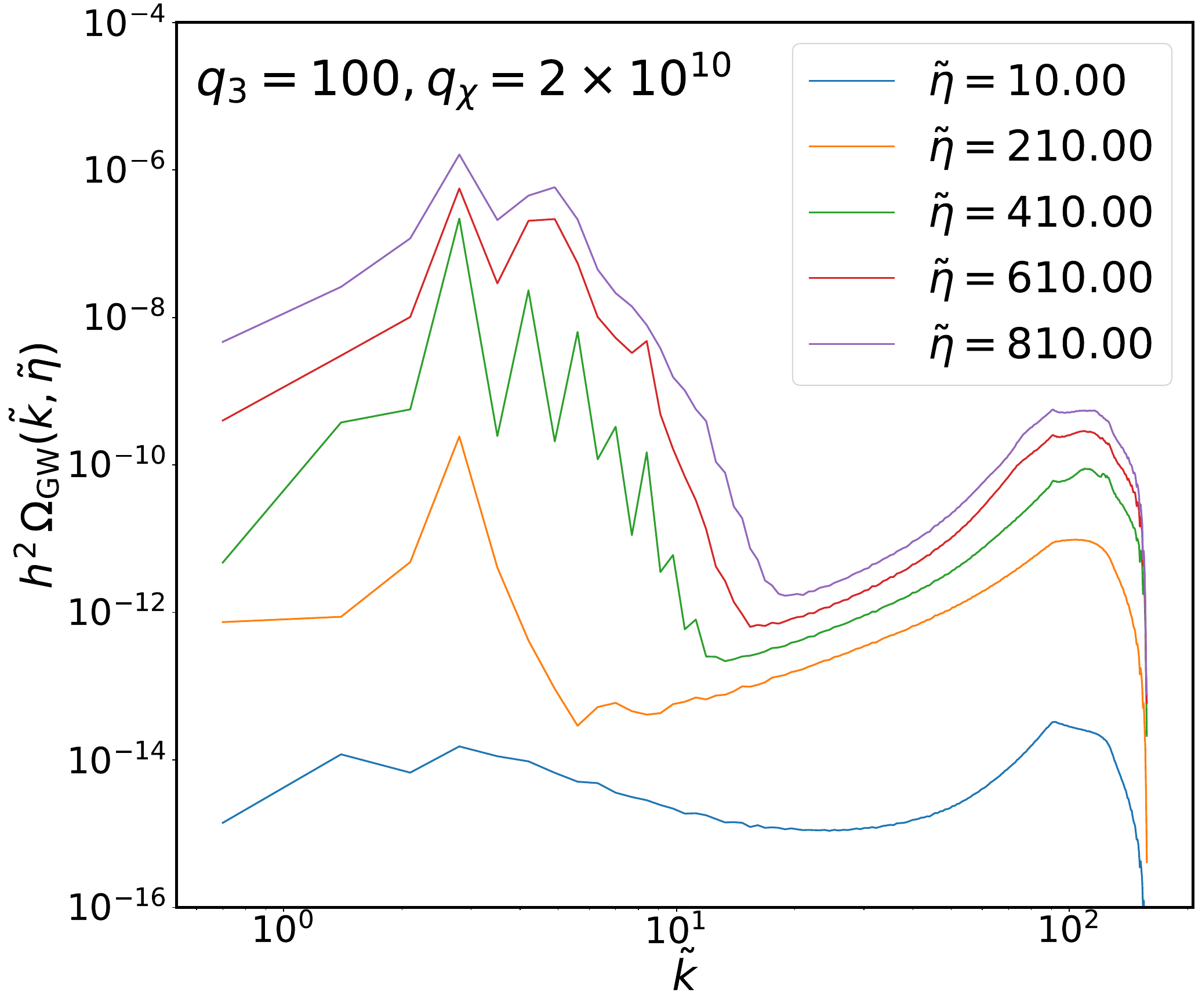}
\hspace{0.2cm}
\includegraphics[width=3.4in, height = 2.8 in]{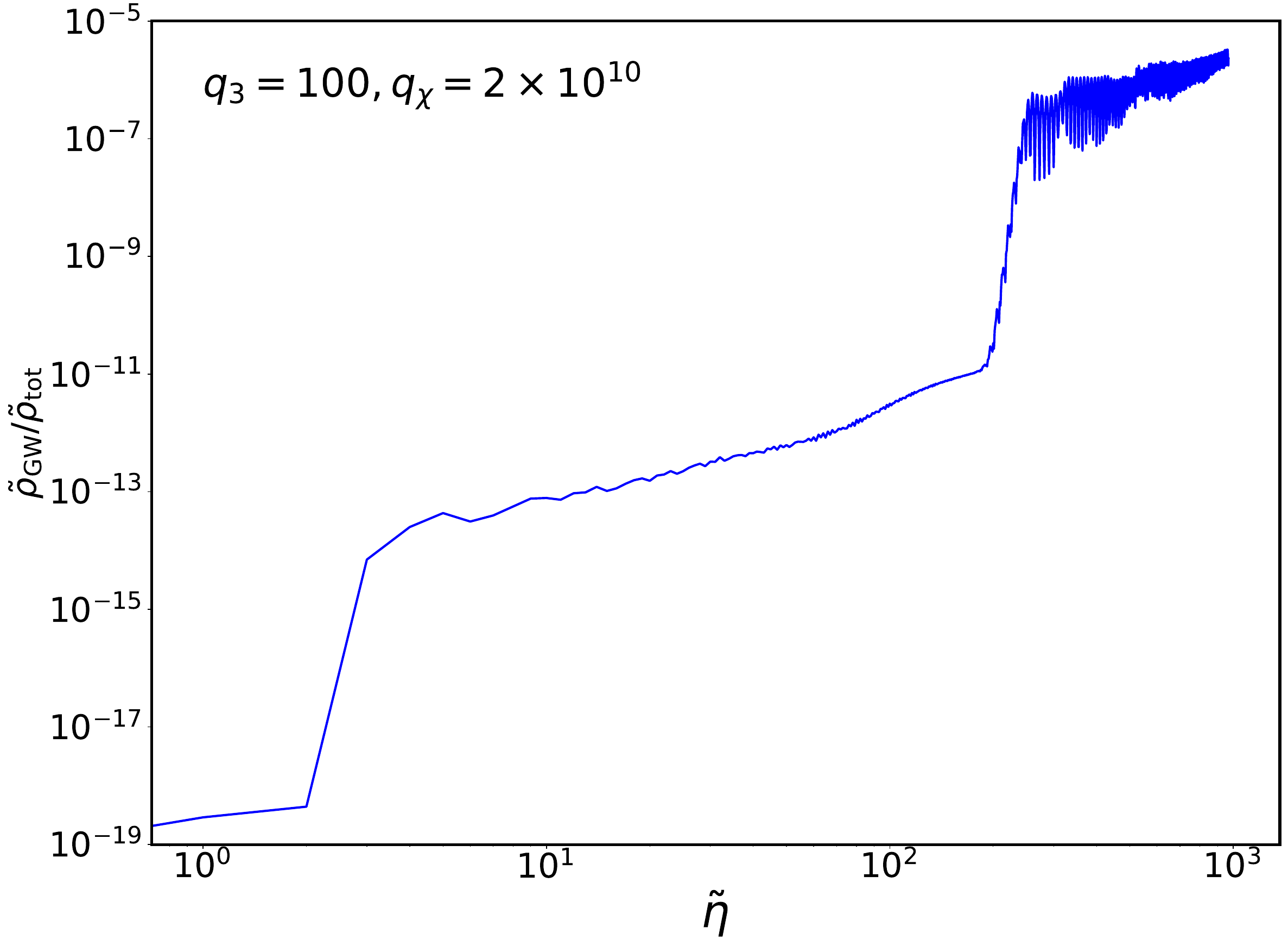}
\vskip -0.1in
\caption{In the left panel, the energy density power spectrum of GWs is shown as a function of the comoving momentum \( \tilde{k} \). In the right panel, the fractional energy density of GWs relative to the total energy density is shown as a function of $\tilde{\eta}$.
}
\label{fig:gw_power_sp}
\end{figure}
In the left panel of Fig.~\ref{fig:gw_power_sp}, we show the power spectrum of GWs at different times, $h^2\Omega_{\rm GW}(\tilde{k}, \tilde{\eta})$, as a function of the comoving wavenumber $\tilde{k}$, produced during the preheating stage. From the figure, we see that as time goes on, the amplitude of the GWs increases until the end of preheating. This has a one-to-one correspondence with the power spectrum of both the inflaton and the daughter fields. The spectrum also exhibits two distinct peaks. The peak at lower frequency corresponds to GWs sourced by parametric resonance, while the peak at higher frequency arises from tachyonic resonance. As discussed earlier, the growth of field fluctuations due to parametric resonance is significantly stronger than that resulting from tachyonic resonance (see Fig.~\ref{fig:variance_plot}). This enhanced amplification of the power spectrum directly translates into a stronger production of GWs. Consequently, the amplitude of the GW spectrum sourced by the inflaton fluctuations, $\delta \tilde{\phi}_{\tilde{k}}$, through parametric resonance is significantly larger than that sourced by the field $\tilde{\chi}_{\tilde{k}}$ through tachyonic resonance. The GW power spectrum reaches its maximum at $\tilde{k} \sim 2$, where the corresponding peak value is $h^2\Omega_{\rm GW}(\tilde{k}, \tilde{\eta}) \sim 10^{-6}$.

In the right panel of Fig.~\ref{fig:gw_power_sp}, we show the time evolution of the fractional energy density of GWs relative to the total energy density of the system. From the figure, we observe that around $\tilde{\eta} \approx 3$, when tachyonic resonance becomes efficient, the energy density of GWs begins to grow. Later, around $\tilde{\eta} \approx 200$, when parametric resonance becomes dominant, a much more rapid growth in the GW energy density is observed. The enhancement in the GW energy density due to parametric resonance is significantly stronger than that produced during the tachyonic phase. Toward the end of preheating, the GW energy density saturates, as the amplification of both the inflaton and the daughter-field fluctuations ceases once the resonance processes terminate.

\begin{figure}[h]
\centering
\includegraphics[width=3.4in, height = 2.8 in]{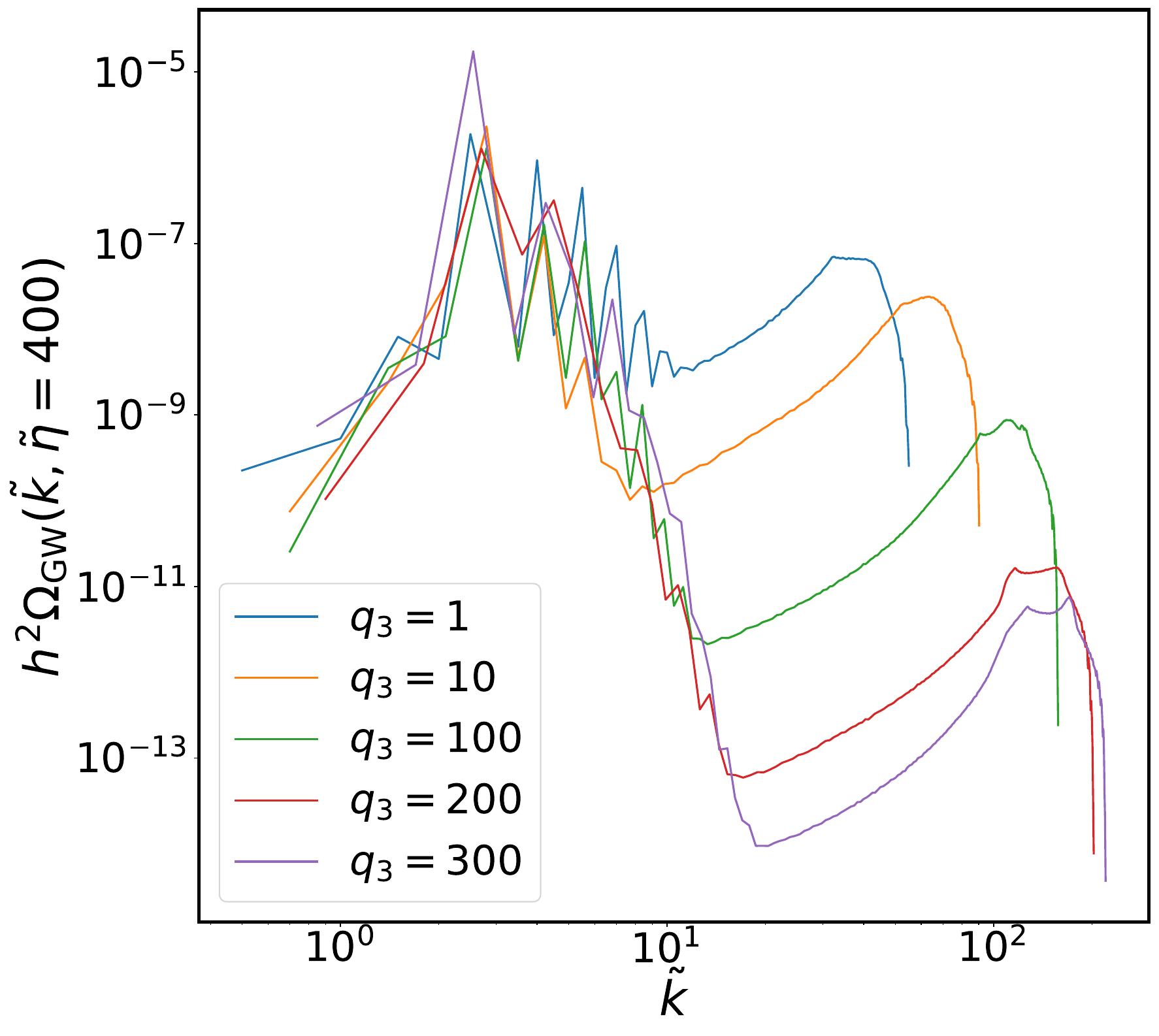}
\hspace{0.2cm}
\includegraphics[width=3.4in, height = 2.8 in]{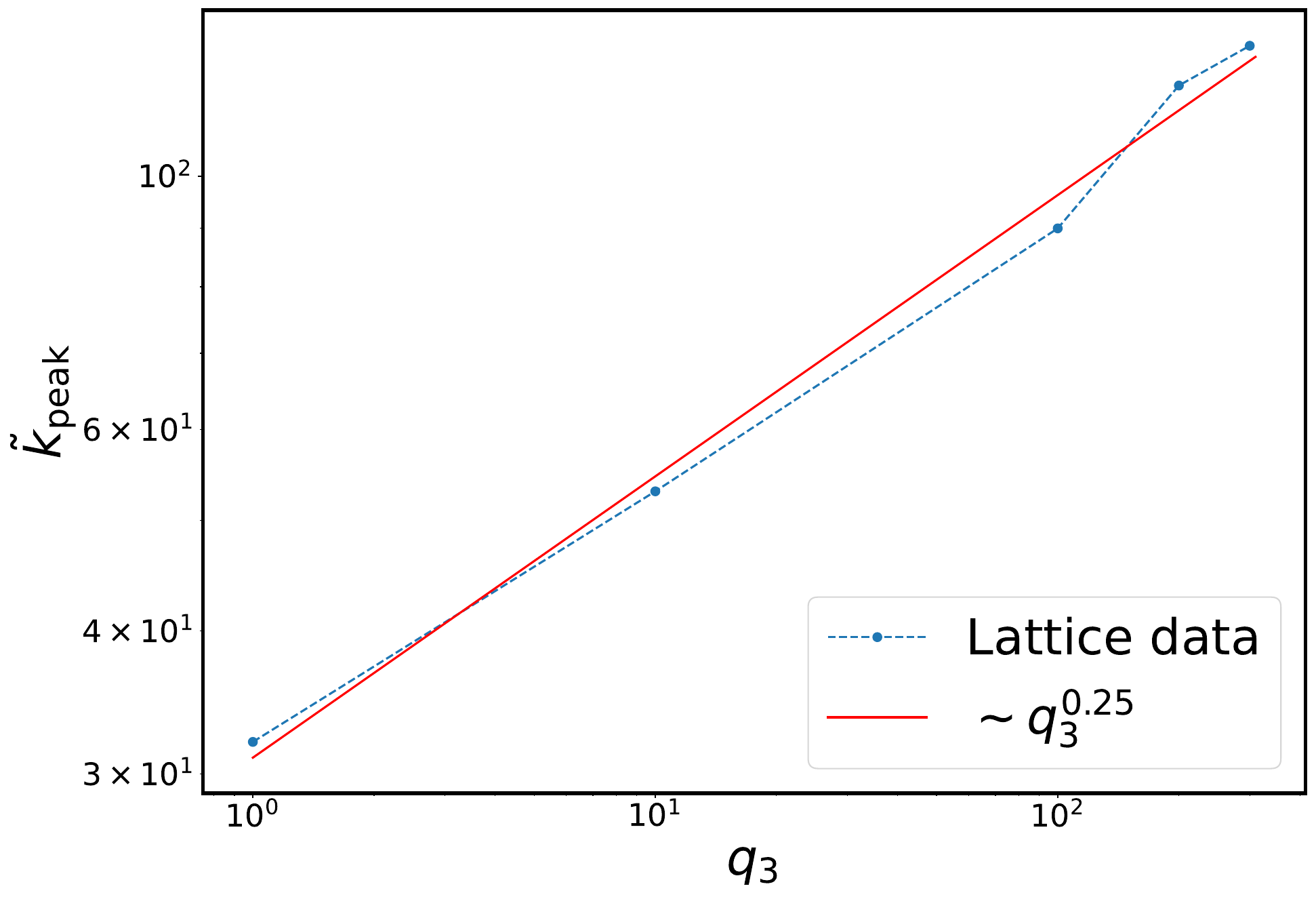}
\vskip -0.1in
\caption{In the left panel, the power spectrum of GWs is shown for different values of the resonance parameter \( q_3 \) at the lattice time \( \tilde{\eta} = 400 \). In the right panel, the peaks of the GW spectra arising from tachyonic resonance for various \( q_3 \) values are shown as blue dots, while the solid red line represents the best-fit function obtained from the lattice data.
}
\label{fig:kvsq3_gw}
\end{figure}

The GWs produced in our model exhibit two distinct peaks, as discussed earlier. Here, we discuss the dependence of these peaks on the resonance parameter $q_3$. In the left panel of Fig.~\ref{fig:kvsq3_gw}, we show the power spectrum of GWs for various values of $q_3$ at $\tilde{\eta}=400$. From the figure, we see that the mode corresponding to the parametric resonance peak remains almost unchanged and is independent of the resonance parameter $q_3$. In contrast, the mode associated with the tachyonic resonance peak significantly depends on $q_3$: the peak shifts towards higher $\tilde{k}$ as $q_3$ increases. This is because increasing $q_3$ enhances the energy transfer to $\tilde{\chi}$, thereby amplifying the production of GWs at higher momentum modes. Due to the strong non-linear nature of the underlying equations, we are unable to determine the relation between $\tilde{k}$ and $q_3$ analytically. However, using lattice simulations, shown in the left panel of Fig.~\ref{fig:kvsq3_gw}, we can empirically approximate this relation.

In the right panel of Fig.~\ref{fig:kvsq3_gw}, we show the mode corresponding to the peak of the tachyonic resonance, $\tilde{k}_{\rm peak}$, as a function of the resonance parameter $q_3$. The blue dots represent the values of $\tilde{k}_{\rm peak}$ obtained from the left panel of Fig.~\ref{fig:kvsq3_gw}, while the red solid line indicates the best-fit curve through these data points. From this fit, we infer the approximate relation $\tilde{k}_{\rm peak} \propto q_3^{0.25}$.

\begin{figure}[H]
    \centering
    \includegraphics[width=4.5in, height = 2.8 in]{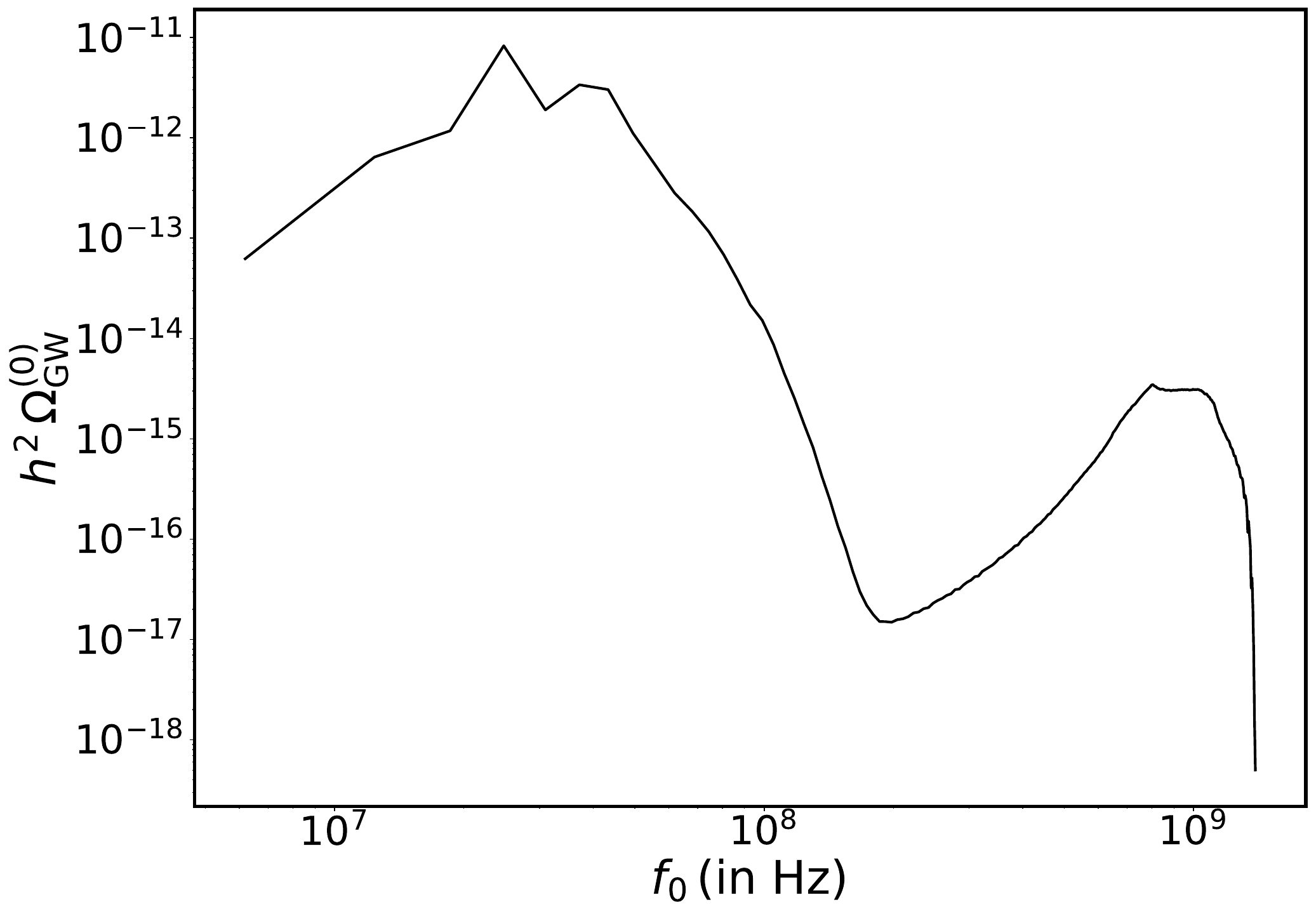}
    \caption{The present-day power spectrum of GWs is shown as a function of frequency.
}
    \label{fig:gw_obs}
\end{figure}
Production of GWs in the context of parametric resonance is well studied in the previous literature~\cite{Easther:2006gt, Figueroa:2017}. In~\cite{Cosme:2023}, the production of GWs in the context of tachyonic resonance is discussed for the $m^2\phi^2$ model. Unlike the case considered in~\cite{Cosme:2023}, we have considered the effects of both parametric and tachyonic resonance, even though in our case, the GW amplitude due to tachyonic resonance is suppressed by four orders of magnitude compared to parametric resonance. Considerable interest in detecting GWs has led to the construction and proposal of various gravitational-wave observatories. These include space-based interferometers such as LISA~\cite{LISAweb}, which probes frequencies in the range $10^{-4}-0.1\,\rm Hz$. The proposed BBO~\cite{Crowder:2005nr} and DECIGO~\cite{Seto:2001qf} missions are sensitive to $0.1-1\,\rm Hz$. Among the ground-based interferometers, the currently operating observatories such as LIGO~\cite{LIGOweb} and VIRGO~\cite{VIRGOweb}, along with the proposed Einstein Telescope (ET)~\cite{ET:2025xjr} and Cosmic Explorer (CE)~\cite{CEweb}, are sensitive up to frequencies of $10^{4}\,\rm Hz$.  

For our case, using Eq.~\eqref{freq_today_final}, the present-day peak frequency of the GW spectrum is found to be $f_{p}^{(0)} \sim 10^{7}\,\mathrm{Hz}$, with a corresponding amplitude of $h^{2}\Omega_{\rm GW}^{(0)} \sim 10^{-11}$ (see Fig.~\ref{fig:gw_obs}). Hence, none of the aforementioned detectors can probe such high-frequency GWs. Among the ultra–high-frequency GW detectors~\cite{Aggarwal:2020olq, Aggarwal:2025noe}, the Holometer~\cite{Holometer:2016qoh} operates in the $1$–$13\,\rm MHz$ frequency range, while the QUEST (Quantum-Enhanced Space-Time) experiment~\cite{Patra:2024eke} works up to $80\,\rm MHz$, with a proposed limit of $200\,\rm MHz$. Therefore, these detectors are, in principle, capable of probing our predicted high-frequency GWs. However, their sensitivity is not yet sufficient to detect GWs with amplitudes as small as $\sim 10^{-11}$.

It is worth mentioning that a clear path toward achieving such sensitivity is currently under development~\cite{Aggarwal:2020olq}. Therefore, we remain hopeful that in the near future, these high-frequency GWs originating from the early universe may also become observable.

\section{Summary and Outlooks}
\label{sec:summary}
In this work, we have studied the dynamics of preheating and the associated production of
gravitational waves (GWs) in an inflationary scenario based on the $\alpha$--attractor potential,
where inflation ends in the positively curved regime of the potential. We considered a trilinear
interaction of the form $h\phi\chi^2$ between the inflaton field $\phi$ and a light scalar
daughter field $\chi$. This coupling naturally gives rise to a complex interplay between \textit{parametric} and \textit{tachyonic} resonance effects, both contributing to the transfer of energy from the inflaton condensate to the inflaton and daughter field fluctuations.

Starting from CMB--consistent parameters of the $\alpha$-attractor potential, we analysed
the post-inflationary evolution using a combination of analytic and numerical methods. In the \textit{linear regime}, we solved the linearised version of inflaton field fluctuation equation and identified the growth rates with Floquet exponent.  The inflaton fluctuations experienced broad-band
parametric amplification, while the daughter field $\chi$ underwent short but intense episodes of
tachyonic growth whenever its effective frequency squared became negative. These complementary
instabilities quickly drove the system away from the homogeneous background configuration.

To capture the full nonlinear evolution and backreaction effects, we performed
three-dimensional lattice simulations using the \texttt{CosmoLattice} framework
\cite{Figueroa:2021,Figueroa:2023}. The simulations revealed that the preheating dynamics
proceed in two distinct stages. In the initial phase, tachyonic resonance in $\chi$ triggers
rapid but transient amplification. Subsequently, parametric resonance in the inflaton field dominates, exciting a certain range of momentum modes and driving a turbulent cascade. The comparison of variances and power spectra shows that the
parametric channel dominates the energy transfer, though the tachyonic phase leaves a
distinct early-time imprint on the dynamics.

The amplified inhomogeneities in both fields generate significant anisotropic stresses, which
act as sources for a stochastic background of gravitational waves. We computed the evolution
of the tensor perturbations and obtained the GW energy spectrum at the end of the
simulation. After redshifting to the present epoch, the spectrum exhibits a
\textit{double-peak structure}: a dominant low-frequency peak associated with the parametric
resonance and a subdominant high-frequency peak originating from the tachyonic bursts. The
present-day peak frequency is found to be $f_p^{(0)} \sim 10^{7}\,\mathrm{Hz}$ with a peak amplitude
$h^2 \Omega_{\mathrm{GW}}^{(0)} \sim 10^{-11}$. Although such ultra-high-frequency GWs are currently beyond the reach of existing detectors such as LIGO, VIRGO, KAGRA, LISA, BBO, or DECIGO, they may be accessible to future MHz-GHz experiments based on
superconducting or optomechanical technologies
\cite{Aggarwal:2020olq, Aggarwal:2025noe}.

Our analysis demonstrates that trilinear preheating in $\alpha$-attractor models naturally
yields a rich multi-channel instability pattern and a distinctive multi-peak GW signal. The
coexistence of parametric and tachyonic resonance provides a robust and general mechanism
for efficient post-inflationary energy transfer. The results also reinforce the role of lattice simulations as an indispensable
tool for exploring the nonlinear regime of early-universe dynamics, where analytic
approximations become unreliable.

Several promising directions remain open for future work:  (a) A detailed study of the later stages of
    reheating and the approach toward full thermal equilibrium could refine the mapping
    between model parameters and inflationary observables~\cite{Antusch:2025ewc}. (b) Including gauge or fermionic sectors
    may introduce additional energy-transfer channels and alter the shape of the resulting GW
    spectrum. (c) Although the predicted GW
    frequencies lie beyond the range of current instruments, future high-frequency GW
    detectors could directly probe this epoch, offering a new observational window on the
    post-inflationary universe. 
    
In conclusion, preheating with trilinear interactions in $\alpha$-attractor inflation provides a
dynamically rich framework in which the nonlinear interplay of tachyonic and parametric
resonances imprints a characteristic multi-scale gravitational-wave signal. These findings
highlight the importance of the post-inflationary epoch not merely as a bridge to the hot Big
Bang, but as a potentially observable source of new cosmological information from the
earliest moments of our universe.

\section*{Acknowledgements}
K.A. acknowledges support from the Axis Bank Post-Doctoral Programme at Ashoka University.

%%%%%%%%%%%%%%%%%%%%%%%%%%%%%%%%%%%%%%%%%%%%%%%%%%%%%%%%%%%%%%%%%%%%%%%%%%%%%%%

%%%%%%%%%%%%%%%%%%%%%%%%%%%%%%%%%%%%%%%%%%%%%%%%%%%%%%%%%%%%%%%%%%%%%%%%%%%%%%%

\bibliographystyle{JHEP}
\bibliography{references}
%%%%%%%%%%%%%%%%%%%%%%%%%%%%%%%%%%%%%%%%%%%%%%%%%%%%%%%%%%%%%%%%%%%%%%%%%%%%%%%
\end{document}